\documentclass[submission,Phys]{SciPost}

\usepackage{xcolor}
\usepackage{amsmath,amssymb}
\usepackage{multirow,hyperref,booktabs,graphicx}
\usepackage{mathtools,feynmp}
\usepackage{multirow}
\usepackage{tikz} 
\usepackage{pgfplots}
   \pgfplotsset{compat=1.5}
   \usepgfplotslibrary{fillbetween}

\graphicspath{{figures/}}

\makeatletter
\@ifundefined{pdfoutput}{}{\DeclareGraphicsRule{*}{mps}{*}{}}
\makeatother

\newcommand{\arxiv}[1]{\href{http://arxiv.org/abs/#1}{arXiv:#1}}

\newcommand{\Dfb}{\mbox{$\raisebox{2mm}{\boldmath ${}^\leftrightarrow$}\hspace{-4mm} D$}}
\newcommand{\Dfba}{\mbox{$\raisebox{2mm}{\boldmath ${}^\leftrightarrow$}\hspace{-4mm} D^a$}}

\newcommand\one{\leavevmode\hbox{\small1\normalsize\kern-.33em1}}

\newcommand{\lag}{\mathcal{L}}

\newcommand{\ope}{\mathcal{O}}
\newcommand{\qqquad}{\qquad \qquad}
\newcommand{\qqqquad}{\qquad \qquad \qquad}

\newcommand{\met}{\slashchar{E}_T}

\newcommand{\gev}{\text{GeV}}
\newcommand{\tev}{\text{TeV}}

\newcommand{\br}{\text{BR}}

\newcommand{\ifb}{\text{fb}^{-1}}

\def\slashchar#1{\setbox0=\hbox{$#1$}           
   \dimen0=\wd0                                 
   \setbox1=\hbox{/} \dimen1=\wd1               
   \ifdim\dimen0>\dimen1                        
      \rlap{\hbox to \dimen0{\hfil/\hfil}}      
      #1                                        
   \else                                        
      \rlap{\hbox to \dimen1{\hfil$#1$\hfil}}   
      /                                         
   \fi}

\newcommand{\eg}{\textsl{e.g.}\;}
\newcommand{\ie}{\textsl{i.e.}\;}

\DeclareMathOperator{\tr}{Tr}

\newcommand{\be}{\begin{eqnarray*}}
\newcommand{\ee}{\end{eqnarray*}}

\newcommand{\bee}{\begin{eqnarray}}
\newcommand{\eee}{\end{eqnarray}}
\newcommand{\beeq}{\begin{equation}}
\newcommand{\eeeq}{\end{equation}}

\begin{document}

\begin{center}{\Large \textbf{
The Gauge-Higgs Legacy of the LHC Run II
}}\end{center}
\begin{center}
Anke Biek\"otter\textsuperscript{1},
Tyler Corbett\textsuperscript{2}, and
Tilman Plehn\textsuperscript{1}
\end{center}

\begin{center}
{\bf 1} Institut f\"ur Theoretische Physik, Universit\"at Heidelberg,
Germany\\ {\bf 2} Niels Bohr International Academy and Discovery
Centre, Niels Bohr Institute, University of Copenhagen, Denmark \\
biekoetter@thphys.uni-heidelberg.de
\end{center}

\begin{center}
\today
\end{center}


\section*{Abstract}
{\bf We present a global analysis of the Higgs and electroweak sector
  based on LHC Run~II and electroweak precision observables. We show
  which measurements provide the leading constraints on Higgs-related
  operators, and how the achieved LHC precision makes it necessary to
  combine rate measurements with electroweak precision
  observables. The SFitter framework allows us to include kinematic
  distributions beyond pre-defined ATLAS and CMS observables,
  independently study correlations, and avoid Gaussian assumptions for
  theory uncertainties. These Run~II results are a step towards a
  precision physics program at the LHC, interpreted in terms of
  effective operators.}

\vspace{10pt}
\noindent\rule{\textwidth}{1pt}
\tableofcontents\thispagestyle{fancy}
\noindent\rule{\textwidth}{1pt}
\vspace{10pt}

\newpage
\begin{fmffile}{feynman}
\section{Introduction}
\label{sec:intro}

After the discovery of a light, likely fundamental Higgs boson largely
compatible with the Standard Model~\cite{higgs}, the LHC has focused
on precision studies of electroweak symmetry
breaking~\cite{review}. From a theoretical as well as from an
experimental perspective, the appropriate interpretation framework for
such LHC precision analyses are effective
Lagrangians~\cite{eft1,eft2,kaoru,kilian,trott}. They require us to
fix the (propagating) particle content and the underlying symmetry
structure.  For the former, experimental observations point to the
Standard Model content, possibly extended by a dark matter agent
coupling to the Higgs sector. Concerning the interactions, we can
assume the Higgs doublet structure of the Standard Model, which
intertwines the Higgs sector and the electroweak gauge
sector~\cite{linear}. The corresponding analyses based on Run~I
data~\cite{legacy1,legacy2,barca,runI_th,runI_ex,laura} and first analyses
based on Run~II data~\cite{runII_eng,runII_concha} prove that the LHC has
successfully transitioned to a precision physics experiment.

In the effective theory version of the Standard Model~\cite{trott} we
assume that departures of Higgs or gauge boson interactions from their
SM predictions are characterized by a new energy scale $\Lambda$. It
is crucial that this energy scale is not kinematically accessible at
the LHC, which means that the corresponding new particles never appear
on their mass shell. This condition defines the validity of the EFT
approach~\cite{validity}. Because the range of energies accessible in the
kinematic regime of the LHC does not guarantee a strong hierarchy of
scales~\cite{englert}, we can then think of an effective Lagrangian
representing classes of new physics
models~\cite{eft_model,higgsmultiplets}.

One of the great advantages of the SMEFT framework is that it allows
for global analyses of LHC measurements not only in the Higgs and
electroweak gauge sectors, but also in the QCD
sector~\cite{qcd_eft,krauss}, the top sector~\cite{top_eft,topfitter},
or the flavor sector~\cite{flavor_eft}. For LHC Run~I there exist
analyses combining Higgs measurements with LEP data~\cite{lep_gauge}
or, even better, di-boson production at the LHC searching for
anomalous triple gauge vertices~\cite{lhc_gauge,legacy2}. At this
point we find that in the effective Lagrangian framework the LHC limits are
surpassing the LEP limits, because effective operators with a momentum
dependence can be tested either through high precision or through
large momentum flow~\cite{lep_lhc}. Similarly, at the level of Run~II
precision we should not hard-code the electroweak precision
constraints into our operator basis~\cite{fermionic}. Fermionic
operators affect electroweak precision data and LHC data in different
combinations with the usual bosonic operators, and this correlation
generally weakens the constraints on operators contributing to Higgs
physics only. This brings the number of SMEFT operators considered in
our global Higgs analysis to 20, plus invisible decays. Two of these
operators turn out to be successfully constrained by non-Higgs
observables, so they do not have to be considered in the actual
analysis.

In this paper we present an \textsc{SFitter} analysis of the Higgs and
gauge sector at the LHC and electroweak precision data.  As usual, we
do not rely on pre-defined results from ATLAS and CMS, but evaluate
event counts in total rate measurements and kinematic distributions
using our in-house framework whenever
available~\cite{sfitter_orig,sfitter_delta}. This allows us to
correlate systematic uncertainties, define our own treatment of
theoretical uncertainties, and account for non-Gaussian
constraints. We start by defining our relevant operator basis in
Sec.~\ref{sec:higgs_sector} and ~\ref{sec:ewpo_sector}. We then
compare possible Higgs-sector constraints on operators measured in
other LHC analyses in Sec.~\ref{sec:top_sector}. With this operator
basis we then report on a global LHC analysis, starting with a
comparison of Run~I and Run~II results, adding electroweak precision
observables, and discussing the interplay of the two kinds of
operators in detail in Sec.~\ref{sec:higgs}. Our final result brings
us a significant step closer to a global \textsc{SFitter} SMEFT
analysis.

\section{Higgs and gauge sector}
\label{sec:higgs_sector}

The linear effective Lagrangian is an $SU(3)_c \otimes SU(2)_L \otimes
U(1)_Y$-symmetric extension of the renormalizable Standard model, but
with the SM field content. It is ordered by inverse
powers of the new physics scale~\cite{eft1,eft2,linear},
\begin{alignat}{5}
\lag = \sum_x \frac{f_x}{\Lambda^2} \; \ope_x \;\;,
\label{eq:def_f}
\end{alignat}
Neglecting lepton number violation at dimension five the first order
of new physics effects is dimension six, with 59 baryon-number
conserving operators, barring flavor structure and Hermitian
conjugation~\cite{eft2}. We follow the definition of the relevant
operator basis of Ref.~\cite{barca}: first, we restrict the initial
set to $P$-even and $C$--even operators\footnote{Before trying to
  prove for example $CP$-violation through a global fit we advocate
  dedicated $CP$ tests for the Higgs and gauge sector~\cite{cp}.}.  We
then use the equations of motion to rotate to a basis where there are
no blind directions linked to electroweak precision data.  We then
neglect all operators that cannot be studied at the LHC yet or which
are strongly constrained from other LHC measurements. This includes
the $HHH$ vertex~\cite{hhh}, the Higgs interactions with
light-generation fermions, and some operators discussed in
Sec.~\ref{sec:top_sector}. We are left with 18 dimension-6
operators, ten of which do not influence electroweak precision
observables at tree level~\cite{barca},
\begin{alignat}{9}
\ope_{GG} &= \phi^\dagger \phi \; G_{\mu\nu}^a G^{a\mu\nu}  \quad 
&\ope_{WW} &= \phi^{\dagger} \hat{W}_{\mu \nu} \hat{W}^{\mu \nu} \phi  \quad 
&\ope_{BB} &= \phi^{\dagger} \hat{B}_{\mu \nu} \hat{B}^{\mu \nu} \phi 
\notag \\
\ope_W &= (D_{\mu} \phi)^{\dagger}  \hat{W}^{\mu \nu}  (D_{\nu} \phi)
& \ope_B &=  (D_{\mu} \phi)^{\dagger}  \hat{B}^{\mu \nu}  (D_{\nu} \phi) 
\notag \\
\ope_{\phi 2} &= \frac{1}{2} \partial^\mu ( \phi^\dagger \phi )
                            \partial_\mu ( \phi^\dagger \phi ) \quad
&\ope_{WWW} &= \tr \left( \hat{W}_{\mu \nu} \hat{W}^{\nu \rho} 
\hat{W}_\rho^\mu \right)  
\label{eq:operators}   \\
\ope_{e\phi,33} &= \phi^\dagger\phi \; \bar L_3 \phi e_{R,3}  \qquad 
&\ope_{u\phi,33} &= \phi^\dagger\phi  \; \bar Q_3 \tilde \phi u_{R,3} \qqquad  
&\ope_{d\phi,33} &= \phi^\dagger\phi \; \bar Q_3 \phi d_{R,3} \notag \; .
\end{alignat}
The covariant derivative acting on the Higgs is $D_\mu = \partial_\mu+
i g' B_\mu/2 + i g \sigma_a W^a_\mu/2$, and the field strengths are
$\hat{B}_{\mu \nu} = i g' B_{\mu \nu}/2$ and $\hat{W}_{\mu\nu} = i
g\sigma^a W^a_{\mu\nu}/2$. This ad-hoc rescaling of the field strength
can be motivated through our expectations from known UV-completions,
but it has no effect on our analysis or its interpretation. The
effective Lagrangian which we use to interpret Higgs and triple-gauge
vertex (TGV) measurements at the LHC is
\begin{align}
\lag_\text{eff} \supset
&- \frac{\alpha_s }{8 \pi} \frac{f_{GG}}{\Lambda^2} \ope_{GG}  
 + \frac{f_{WW}}{\Lambda^2} \ope_{WW} 
 + \frac{f_{BB}}{\Lambda^2} \ope_{BB} \notag \\
&+ \frac{f_W}{\Lambda^2} \ope_W  
 + \frac{f_B}{\Lambda^2} \ope_B  
 + \frac{f_{\phi 2}}{\Lambda^2} \ope_{\phi 2}
 + \frac{f_{WWW}}{\Lambda^2} \ope_{WWW} \notag \\
&+ \frac{f_\tau m_\tau}{v \Lambda^2} \ope_{e\phi,33} 
 + \frac{f_b m_b}{v \Lambda^2} \ope_{d\phi,33} 
 + \frac{f_t m_t}{v \Lambda^2} \ope_{u\phi,33}
 + \text{invisible decays}\;.
\label{eq:ourlag1}
\end{align}
For invisible Higgs decays we do not include a term in the Lagrangian
and consequently describe it in terms of an invisible partial width.
It is best constrained through WBF Higgs production~\cite{eboli_zeppenfeld}.
All operators except for $\ope_{WWW}$ contribute to Higgs
interactions.  Their contributions to the several Higgs vertices,
including non-SM Lorentz structures, are described in
Ref.~\cite{legacy1,deBlas:2018tjm}.\medskip
 
Some of the operators in Eq.~\eqref{eq:operators} contribute to the
self-interactions of the electroweak gauge bosons. They can be linked
to specific deviations in the Lorentz structures entering the $WWZ$
and $WW\gamma$ interactions, historically written as $\kappa_\gamma,
\kappa_Z, g_1^Z, g_1^\gamma$, $\lambda_\gamma$, and
$\lambda_Z$~\cite{kaoru_tgv}. After using electromagnetic gauge
invariance to fix $g_1^\gamma = 1$, the shifts are defined by
\begin{align}
\Delta \lag_\text{TGV} =& 
- i e \; (\kappa_\gamma-1) \; W^+_\mu W^-_\nu \gamma^{\mu \nu}
- \frac{i e \lambda_\gamma}{ m_W^2} \; W_{\mu \nu}^+ W^{- \nu \rho} \gamma_\rho^\mu
- \frac{i g_Z \lambda_Z}{ m_W^2} \; W_{\mu \nu}^+ W^{- \nu \rho} Z_\rho^{\;\mu} 
\notag \\
&- i g_Z \; (\kappa_Z -1) \; W^+_\mu W^-_\nu Z^{\mu \nu}
- i g_Z \; ( g_1^Z -1) \; \left( W^+_{\mu \nu} W^{- \mu} Z^\nu - W^+_\mu Z_\nu W^{- \mu \nu} 
                    \right) \notag \\
=& - i e \; \frac{g^2 v^2}{8 \Lambda^2} \left( f_W + f_B \right)  \; W^+_\mu W^-_\nu \gamma^{\mu \nu}
- i e \; \frac{3 g^2 f_{WWW}}{4 \Lambda^2} \; W_{\mu \nu}^+ W^{- \nu \rho} \gamma_\rho^\mu \notag \\
&- i g_Z \; \frac{g^2 v^2}{8 c_w^2 \Lambda^2} \left( c_w^2 f_W - s_w^2 f_B \right) \; W^+_\mu W^-_\nu Z^{\mu \nu}
- i g_Z \; \frac{3 g^2 f_{WWW}}{4 \Lambda^2} \; W_{\mu \nu}^+ W^{- \nu \rho} Z_\rho^{\; \mu} \notag \\
&- i g_Z \; \frac{g^2 v^2 f_W}{8 c_w^2 \Lambda^2} \; \left( W^+_{\mu \nu} W^{- \mu} Z^\nu - W^+_\mu Z_\nu W^{- \mu \nu} 
                    \right) \; ,
\label{eq:tgvlag}
\end{align}
where $e = g s_w$ and $g_Z = g c_w$. The two notational conventions
are equivalent for gauge-invariant models and linked as
\begin{align}
\kappa_\gamma &= 1+ 
 \frac{g^2 v^2}{8\Lambda^2}
\left( f_W + f_B \right) \qqqquad 
\kappa_Z = 1+ \frac{g^2 v^2}{8 c_w^2\Lambda^2} \left(c_w^2 f_W - s_w^2 f_B \right) \notag \\  
g_1^Z &= 1+ \frac{g^2 v^2}{8 c_w^2\Lambda^2} f_W \qqqquad
g_1^\gamma = 1 \qqqquad  
\lambda_\gamma = \lambda_Z = 
\frac{3 g^2 m_W^2}{2 \Lambda^2} f_{WWW}\; .
\label{eq:wwv}
\end{align}
The three Wilson coefficients relevant for our analysis of di-boson
production are $f_B$, $f_W$ and $f_{WWW}$, plus the operators
influencing electroweak precision data discussed in Section~\ref{sec:ewpo_sector}. To get a
very rough idea what kind of new physics scales we can probe in the
electroweak gauge and Higgs sector we quote the typical range from the
global Run~I analyses,
\begin{align}
\frac{\Lambda}{\sqrt{|f|}} \gtrsim 300~...~500~\gev
\qquad \text{(Higgs-gauge analysis at Run~I~\cite{legacy2}).}
\label{eq:rough_higgs}
\end{align}
We note that already the Run~I di-boson measurements clearly
outperform the corresponding LEP measurements evaluated in the
effective operator basis of Eq.\eqref{eq:ourlag1}.\medskip

If we deviate from this scenario and consider instead the more
generic non-linear or chiral effective
Lagrangian~\cite{nonlinear1,nonlinear2}, the parametrization would be
extended. In the most generic scenario, the TGV couplings defined
above depend on a larger number of parameters and the correlations
from gauge dependence are lost. Furthermore, the deviations generated
by non-linear operators in the TGVs could be completely de-correlated
to the deviations generated in the Higgs interactions. For the Higgs
sector alone, the linear and non-linear analyses can be trivially
mapped onto each other~\cite{legacy1}.

\section{Electroweak precision sector}
\label{sec:ewpo_sector}

While the Lagrangian in Eq.\eqref{eq:ourlag1} does not include
tree-level contributions to electroweak precision observables, we know
that at the level of 13~TeV data the corresponding operators should
not be neglected~\cite{fermionic,runII_eng,runII_concha}. This means that we need
to add two bosonic operators 
\begin{align}
\ope_{\phi 1} = (D_\mu \phi)^\dagger \; \phi \phi^\dagger \; (D^\mu \phi)
\qqqquad 
\ope_{BW} = \phi^\dagger \hat{B}_{\mu\nu} \hat{W}^{\mu\nu} \phi \; ,
\label{eq:ope_new1}
\end{align}
which affect gauge and Higgs interactions. In addition we consider the
fermionic Higgs-gauge operators
\begin{align}
\ope_{\phi Q,ij}^{(1)} &=\phi^\dagger (i\,{\Dfb}_{\mu} \phi) (\bar Q_{i}\gamma^\mu Q_i) \qquad & \ope_{\phi Q,ij}^{(3)} &=\phi^\dagger (i\,{\Dfba}_{\!\!\mu} \phi) (\bar Q_i\gamma^\mu \frac{\sigma_a}{2} Q_i)  \notag \\
\ope_{\phi L,ij}^{(1)} &=\phi^\dagger (i\,{\Dfb}_{\mu} \phi) (\bar L_{i}\gamma^\mu L_i) \qquad & \ope_{\phi L,ij}^{(3)}&=\phi^\dagger (i\,{\Dfba}_{\!\!\mu} \phi) (\bar L_i\gamma^\mu \frac{\sigma_a}{2} L_i)  \notag \\
\ope_{\phi u,ij}^{(1)} &=\phi^\dagger (i\,{\Dfb}_{\mu} \phi) (\bar u_{R,i}\gamma^\mu u_{R,i}) \qquad & \ope_{LLLL} &= (\bar{L}_1 \gamma_\mu L_2) \; (\bar{L}_2 \gamma^\mu L_1) \notag \\
\ope_{\phi d,ij}^{(1)} &=\phi^\dagger (i\,{\Dfb}_{\mu} \phi) (\bar d_{R,i}\gamma^\mu d_{R,i}) \notag \\
\ope_{\phi e,ij}^{(1)}&=\phi^\dagger (i\,{\Dfb}_{\mu} \phi) (\bar e_{R,i}\gamma^\mu e_{R,i}) \notag \\
\ope_{\phi ud,ij}^{(1)} &=\tilde \phi^\dagger (i\,{\Dfb}_{\mu} \phi) (\bar u_{R,i}\gamma^\mu d_{R,i}) 
\label{eq:ope_new2}
\end{align}
The operator $\ope_{\phi ud,ij}^{(1)}$ contains the charged current $\bar{u}_R
\gamma^\mu d_R$~\cite{Cirigliano:2009wk,Falkowski:2017pss,Alioli:2017ces,runII_concha}. Given that it does not
interfere with the Standard Model and the known flavor physics
constraints we will ignore it in our analysis, the same way we exclude
for example dipole operators.  

\begin{table}[b!]
\begin{center}
\begin{tabular}{l | ccccc}
\toprule
operator & $H f \bar f$ & $Z q q$ & $W q q'$ & $Z l \bar l$ & $W l \nu$  \\ 
\midrule
 $\ope_{\phi 1}$     &    $\times$   &  $\times$ &  $\times$ & $\times$ & $\times$\\
 $\ope_{BW}$     &           &  $\times$ & $\times$ & $\times$ & $\times$\\
 $\ope_{\phi Q}^{(3)}$     &    & $\times$  & $\times$ \\
 $\ope_{\phi Q}^{(1)}$     &    & $\times$  &  \\
 $\ope_{\phi u}^{(1)}$     &    & $\times$ &  \\
 $\ope_{\phi d}^{(1)}$     &    & $\times$ &  \\
 $\ope_{\phi e}^{(1)}$     &    &   & & $\times$ \\
 \bottomrule
\end{tabular}
\end{center}
\caption{List of operators affecting electroweak precision observables
  and their effect on fermionic couplings testable at the LHC.}
\label{tab:operator_influence}
\end{table}

The first eight operators generate anomalous weak boson couplings to
fermions, while they do not affect the Higgs coupling to fermions, see
Tab.~\ref{tab:operator_influence}.  They do modify the Higgs couplings
to weak bosons and fermions, for instance introducing point-like
$HVff$ interactions. We also include the 4-lepton operator
$\ope_{LLLL}$ as it induces a shift in the Fermi constant. For our
study we assume diagonal and generation independent Wilson
coefficients for the fermionic operators affecting the electroweak
currents.  Further, we will eliminate the leptonic operators
$\ope_{\phi L,ii}^{(1)}$ and $\ope_{\phi L,ii}^{(3)}$ using the
equations of motion:
\begin{align}
&2 \ope_{\phi 2}+ 2 \ope_{\phi 4} =\sum_{ij}\left(
	y_{ij}^e (\ope_{e\phi,ij})^\dagger + y_{ij}^u \ope_{u\phi,ij}  + y_{ij}^d (\ope_{d\phi,ij})^\dagger 
	+ \text{h.c.}\right)
	- \frac{\partial V(h)}{\partial h}
\notag \\
&2\ope_B+\ope_{BW}+\ope_{WW}+g^2\left(\ope_{\phi 4}-\frac{1}{2}\ope_{\phi 2}\right) =-\frac{g^2}{4}\sum_i\left(\ope_{\phi L,ii}^{(3)}+\ope_{\phi Q,ii}^{(3)}\right)
\notag \\
&2\ope_B+\ope_{BW}+\ope_{BB}+g'^2 \left(\ope_{\phi 1}-\frac{1}{2}\ope_{\phi 2}\right) = \notag \\
&\phantom{Halllllloooooooooo} -\frac{g'^2}{2}\sum_i\left(-\frac{1}{2}\ope_{\phi L,ii}^{(1)}+\frac{1}{6}\ope_{\phi Q,ii}^{(1)}-\ope_{\phi e,ii}^{(1)}+\frac{2}{3}\ope_{\phi u,ii}^{(1)}-\frac{1}{3}\ope_{\phi d,ii}^{(1)}\right) \; .
\end{align}
Assuming a universal flavor structure 
this leaves us with the additional contributions to our effective
Lagrangian,
\begin{align}
\lag_\text{eff} \supset
&+ \frac{f_{\phi 1}}{\Lambda^2} \ope_{\phi 1} 
 + \frac{f_{BW}}{\Lambda^2} \ope_{BW} 
 + \frac{f_{LLLL}}{\Lambda^2} \ope_{LLLL} \notag \\
&+ \frac{f_{\phi Q}^{(1)}}{\Lambda^2} \ope_{\phi Q}^{(1)} 
 + \frac{f_{\phi d}^{(1)}}{\Lambda^2} \ope_{\phi d}^{(1)} 
 + \frac{f_{\phi u}^{(1)}}{\Lambda^2} \ope_{\phi u}^{(1)} 
 + \frac{f_{\phi e}^{(1)}}{\Lambda^2} \ope_{\phi e}^{(1)} 
+ \frac{f_{\phi Q}^{(3)}}{\Lambda^2} \ope_{\phi Q}^{(3)} \; . 
\label{eq:ourlag2}
\end{align}
Together with Eq.\eqref{eq:ourlag1} this defines the operator basis
for our global analysis, altogether 18 operators plus the invisible
Higgs branching ratio.  While the additional operators affect many of
our LHC measurements, they are also strongly constrained by
electroweak precision observables. The challenge is that the bosonic
operators in Eq.\eqref{eq:ope_new1} and the fermionic operators in
Eq.\eqref{eq:ope_new2} not only contribute to electroweak precision
physics, but also to di-boson or Higgs production at an observable
level, where they are included e.g.~in our study of triple gauge vertices. 
Because the two data sets combine very different combinations
of operators, we have to combine our Run~II analysis with a set of
electroweak precision observables.  We follow Ref.~\cite{unitarity}
and review this approach briefly.  Our $Z$-pole observables are
\begin{align}
\Big\{ \Gamma_Z,\, \sigma_h^{0},\, 
   \mathcal{A}_l(\tau^{\rm pol}),\, 
   R_l^0,\, \mathcal{A}_l(\text{SLD}),\, A_\text{FB}^{0,l},\, 
   R_c^0,\, R_b^0,\, \mathcal{A}_c,\, \mathcal{A}_b,\, A_\text{FB}^{0,c},\, A_\text{FB}^{0,b}(\text{SLD/LEP-I}) \Big\} \; .
\label{eq:zobs}
\end{align}
with measurements and correlations taken from
Ref~\cite{ewwg}. We also include the $W$-observables
\begin{align}
\Big\{ m_W,\, \Gamma_W,\, \br (W\to l\nu) \Big\} \;  ,
\label{eq:wobs}
\end{align}
with values taken from Ref.~\cite{pdg}.  The SM predictions for these
observables are taken from Ref.~\cite{deBlas:2017wmn}. We note that
for the SM prediction of the $W$-mass this includes the full one- and
two-loop EW and two-loop QCD corrections of
$\mathcal{O}(\alpha\alpha_s)$ as well as some 3-loop
contributions.  The contributions of our dimension-6 operators can be
found in Ref.~\cite{unitarity}, where we limit ourselves to linear
contributions from the higher-dimensional operators considered in our
fit. This approximation is justified as long as the dimension-6
corrections are small, \ie $f m_Z^2/\Lambda^2 \ll 1$ assuming that the
typical energy scale of electroweak precision data is around
$m_Z$. The standard analyses of electroweak precision data indeed give
individual limits of the kind
\begin{align}
\frac{\Lambda}{\sqrt{|f|}} \gtrsim 4~...~10~\tev
\qquad \text{(electroweak precision data~\cite{deBlas:2017wmn}).}
\label{eq:rough_ewpo}
\end{align}
These limits significantly exceeds the expected sensitivity of the
global LHC analysis from Eq.\eqref{eq:rough_higgs}, which naively
suggests that it is not necessary to combine the two sectors. In the
discussion of our global fit in Section~\ref{sec:higgs} we will see
how the fermionic Higgs-gauge operators nevertheless lead to visible
effects at the LHC.

\section{QCD and top sectors}
\label{sec:top_sector}

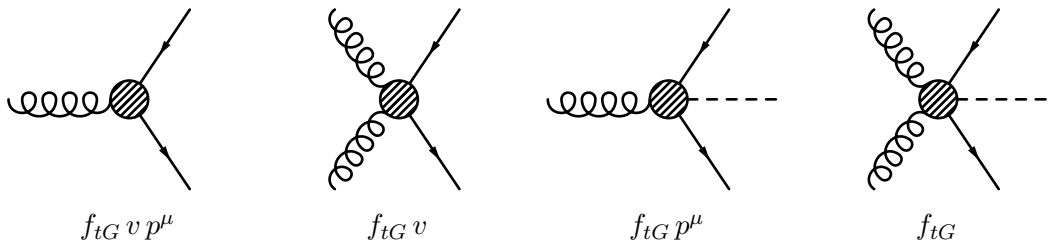
\begin{figure}[b!]
\begin{tabular}{cccc}
\begin{fmfgraph*}(90,75)
\fmfset{arrow_len}{2mm}
\fmfsurroundn{v}{12}
\fmf{gluon}{v7,o1}
\fmf{fermion}{v3,o1,v11}
\fmfv{decoration.shape=circle,decoration.filled=shaded}{o1}
\end{fmfgraph*}
&\begin{fmfgraph*}(90,75)
\fmfset{arrow_len}{2mm}
\fmfsurroundn{v}{12}
\fmf{phantom}{v7,o1}
\fmf{gluon,tension=0}{v5,o1,v9}
\fmf{fermion}{v3,o1,v11}
\fmfv{decoration.shape=circle,decoration.filled=shaded}{o1}
\end{fmfgraph*}
&\begin{fmfgraph*}(90,75)
\fmfset{arrow_len}{2mm}
\fmfsurroundn{v}{12}
\fmf{gluon}{v7,o1}
\fmf{fermion}{v3,o1,v11}
\fmf{dashes,tension=0}{o1,v1}
\fmfv{decoration.shape=circle,decoration.filled=shaded}{o1}
\end{fmfgraph*}
&\begin{fmfgraph*}(90,75)
\fmfset{arrow_len}{2mm}
\fmfsurroundn{v}{12}
\fmf{phantom}{v7,o1}
\fmf{gluon,tension=0}{v5,o1,v9}
\fmf{fermion}{v3,o1,v11}
\fmf{dashes,tension=0}{o1,v1}
\fmfv{decoration.shape=circle,decoration.filled=shaded}{o1}
\end{fmfgraph*} \\
 $f_{tG} \, v \, p^\mu$
&$f_{tG} \, v$
&$f_{tG} \, p^\mu$
&$f_{tG}$
\end{tabular}
\caption{Interactions through the chromo-magnetic top operator.  The
  vertices scaling with $p^\mu$ come from the derivative in the field
  strength, while those scaling with $v$ are generated by the
  commutator component of the field strength.}
\label{fig:OtGdiagrams}
\end{figure}

An operator which should be added to any basis confronted with LHC
data is the anomalous triple gluon coupling
\begin{align}
\ope_G &= f_{abc} 
         G_{a \nu}^\rho G_{b \lambda}^\nu G_{c \rho}^\lambda 
\qquad \text{with} \qquad 
\lag_\text{eff} \supset 
\frac{g_s f_G}{\Lambda^2} \, \ope_G \; ,
\label{eq:g3}
\end{align}
with $G_a^{\rho \nu} = \partial^\rho G_a^\nu - \partial^\nu G_a^\rho -
i g_s f_{abc} G^{b \rho} G^{c \nu}$. It contributes to any
gluon-induced LHC process, for instance Higgs production with a hard
jet. While it only affects kinematic distributions with an additional
hard parton, it needs to be taken into account when we use the same
distribution to separate $\ope_{u \phi,33}$ effects from $\ope_{GG}$.
On the other hand, it can be constrained by ATLAS multi-jet data at
13~TeV, giving the 95\% CL limits~\cite{krauss}
\begin{align}
\frac{\Lambda}{\sqrt{f_G}} > 5.2 \, (5.8)~\tev 
\qquad \text{observed (expected) from multi-jets.}
\label{eq:reach_qcd}
\end{align}
This limits the possible effects on Higgs production rates beyond
anything a global Higgs analysis would be sensitive to in the absence
of a dedicated enhancement mechanism in Higgs rates.\medskip

A critical feature of Higgs analyses is the combination of direct and
indirect measurements of the top-Higgs coupling in gluon fusion and
associated Higgs-top production~\cite{eft_higgs_top}. The
chromo-magnetic top operator
\begin{align}
\ope_{tG}=(\bar Q\sigma^{\mu\nu}T^Au_R) \; \tilde H \; G_{\mu\nu}^A \; ,
\end{align}
will, in principle, affect these observables~\cite{runII_eng} and has
been studied extensively in top-EFT analyses~\cite{topfitter}. The
interaction vertices induced by $\ope_{tG}$ are shown in
Fig.~\ref{fig:OtGdiagrams}. The first two diagrams contribute to top
pair production, the second set to $t\bar t H$ production. In each
case one of the interactions is proportional to the momentum flowing
through the vertex.\medskip

To constrain $f_{tG}$ in a Higgs fit we can consider gluon fusion and
$t\bar{t}H$ production with additional jets. However, extra hard
gluons in the final state are a typical higher-order effect and likely
suppressed.  Alternatively, we can use momentum-dependent
distributions in $t\bar{t}H$ production.  The third vertex in
Fig.~\ref{fig:OtGdiagrams} appears to allow for such effects as it only includes
a single gluon, however this momentum dependence as well as the triple gluon 
vertex of the SM will be compensated by an additional propagator in the 
amplitude resulting in no additional growth with momentum. This lack of 
growth with momentum is demonstrated in Figure~\ref{fig:HTdistribution} below which shows the shape
of the $H_T$ distribution does not change dramatically with increasing $f_{tG}$.

We can estimate the extent to which this operator can be
constrained. The most promising distribution currently available is
the $H_T$ distribution in the all-hadronic
\begin{align}
pp \to t\bar t H \to t \bar{t} \; b \bar{b}
\end{align}
signature released by CMS~\cite{Sirunyan:2018ygk}. In
Fig.~\ref{fig:HTdistribution} we reproduce their $H_T$ distribution as
well as the distribution in the presence of two benchmark values of
$f_{tG}$.  We generate the relevant $ttH$ process merged with one
additional jet using \textsc{Madgraph5}~\cite{madgraph} and
\textsc{Pythia8}~\cite{pythia}, combined with
\textsc{Delphes3}~\cite{delphes}.  The two benchmarks each correspond
to
\begin{align}
\frac{\Lambda}{\sqrt{|f_{tG}|}} &\gtrsim 1~\tev
&\qquad &\text{(top sector~\cite{topfitter})}\notag \\
\frac{\Lambda}{\sqrt{|f_{tG}|}} &\gtrsim 320~\gev
&\qquad &\text{(Higgs sector~\cite{runII_eng})}
\label{eq:reach_top}
\end{align}
From Fig.~\ref{fig:HTdistribution} we see that our expected
sensitivity is comparable to the Higgs study and not competitive with
the top-sector constraints.  This comes as no
surprise: the $t\bar t H$ cross section is phase space suppressed relative
to $t\bar t$ production and its cross section at the LHC is measured 
to be approximately three orders of magnitude below that of $t\bar t$ 
production~\cite{ATLAS13tthObs,ATLASttXS}.
In addition, $t\bar{t}H$
production is plagued by large backgrounds. This implies statistical limitations on measuring 
$f_{tG}$ in $t\bar t H$, so indeed $\ope_{tG}$ and
$\ope_G$ can both be neglected in global Higgs analyses in the near future. 
There are projections, however, that for the 14 TeV LHC with a luminosity of 3/ab 
that constraints on $\ope_{tG}$ from $t\bar t H$ will exceed those of $t\bar t$ 
\cite{ttheventuallybeatstt}.

\pgfkeys{/pgf/number format/.cd,1000 sep={\,}}
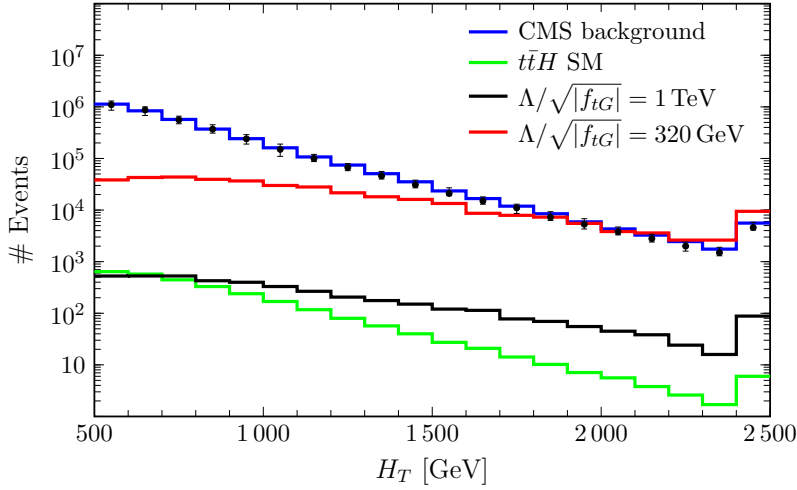
\begin{figure}[t]
\centering
\resizebox{0.7\textwidth}{!}{
\begin{tikzpicture} \begin{semilogyaxis}
[
    axis line style = thick,
    minor tick num = 4,
    ymin = 1,ymax = 100000000,
    xmin = 500,xmax = 2500,
    xlabel={\large $H_T$ [GeV]},
    xtick = {500,1000,1500,2000,2500},
    every x tick/.style={color=black, thin},
        every y tick/.style={color=black, thin},
   ylabel={\large $\#$ Events},
    yticklabels = {$$,$10$,$10^2$,$10^3$,$10^4$,$10^5$,$10^6$,$10^7$},
    tick label style={font=\normalsize},
     legend style={
        cells={anchor=west},
        draw=none,
        column sep=2pt,
    },
    legend columns=1,
    legend entries={CMS background,$t\bar{t}H$ SM,$\Lambda/\sqrt{|f_{tG}|}= 1\, \tev$,$\Lambda/\sqrt{|f_{tG}|} = 320\, \gev$},
    width=1.51*8cm,
    height=8cm
        ]

    \addplot [
        const plot, draw=blue, line width=0.05cm
    ] coordinates {
        (500,1130000)  	(600,838000)
        (700,571000)    	(800,371000)
        (900,242000)    	(1000,161000)
        (1100,107000)  	(1200,74500)
        (1300,50700)    	(1400,35300)
        (1500,23500)    	(1600,16700)
        (1700,11900)    	(1800,8490)
        (1900,5910)    	(2000,4310)
        (2100,3270)    	(2200,2450)
        (2300,1750)    	(2400,5590) (2500,5590)
    };
    
        \addplot [
        const plot, draw=green, line width=0.05cm
    ] coordinates {
        (500,640.1)	(600,573.9)
        (700,444.3)	(800,329.3)
        (900,239.4)	(1000,169.0)
        (1100,117.3)	(1200,79.8)
        (1300,56.9	)	(1400,40.0)
        (1500,27.3	)	(1600,20.9)
        (1700,14.2)	(1800,10.2)
        (1900,7.1)		(2000,5.6)
        (2100,3.8)		(2200,2.6)
        (2300,1.7)		(2400,6.0)(2500,6.0)
        };
        
            \addplot [
        const plot, draw=black, line width=0.05cm
    ] coordinates {
        (500,526.8)	(600,527.9)
        (700,527.9)	(800,426.7)
        (900,398.8)	(1000,329.9)
        (1100,266.4)	(1200,205.7)
        (1300,176.2)	(1400,150.4)
        (1500,120.4)	(1600,113.8)
        (1700,77.7)	(1800,69.5)
        (1900,55.3)	(2000,44.9)
        (2100,38.3)	(2200,24.1)
        (2300,15.9)	(2400,88.1)(2500,88.1)
        };
          \addplot [
        const plot, draw=red, line width=0.05cm
    ] coordinates {
        (500,38388.2)	(600,42783.5)
        (700,43671.9)	(800,39370.2)
        (900,36705.0)	(1000,29971.8)
        (1100,28008.0)	(1200,21602.2)
        (1300,18095.4)	(1400,16084.8)
        (1500,13419.6)	(1600,8650.2)
        (1700,7902.2)	(1800,7294.3)
        (1900,5517.5)	(2000,3834.2)
        (2100,3600.4)	(2200,2618.5)
        (2300,2618.5)	(2400,9445.2)(2500,9445.2)
        };
        \addplot+ [
        only marks,
        black,
        mark options={black},
        mark=*,
        mark size=1.3pt,
        error bars/.cd,
        y dir=both, y explicit,
    ] coordinates {
        (550,1100000) += (0,200000) -= (0,240000)
        (650,870000) += (0,130000) -= (0,190000)
        (750,560000) += (0,100000) -= (0,90000)
        (850,370000) += (0,80000) -= (0,60000)
        (950,240000) += (0,50000) -= (0,50000)
        (1050,150000) += (0,40000) -= (0,40000)
        (1150,100000) += (0,20000) -= (0,13000)
        (1250,67000) += (0,13000) -= (0,9000)
        (1350,46000) += (0,10000) -= (0,6000)
        (1450,31000) += (0,7000) -= (0,4000)
        (1550,21000) += (0,6000) -= (0,2000)
        (1650,15000) += (0,3000) -= (0,2000)
        (1750,11000) += (0,2000) -= (0,2400)
        (1850,7300) += (0,2000) -= (0,1000)
        (1950,5300) += (0,1500) -= (0,1000)
        (2050,3800) += (0,900) -= (0,500)
        (2150,2800) += (0,600) -= (0,400)
        (2250,2000) += (0,400) -= (0,400)
        (2350,1500) += (0,400) -= (0,200)
        (2450,4600) += (0,1200) -= (0,500)
    };
\end{semilogyaxis}
\end{tikzpicture}
}
\caption{$H_T$ distributions for $t\bar{t}H$ production for the
  Standard Model, $\Lambda/\sqrt{|f_{tG}|} = 1$~TeV corresponding to
  the top physics limit, and $\Lambda/\sqrt{|f_{tG}|} = 320$~GeV
  corresponding to the Higgs physics limit. The background estimate
  and the data points are from Ref.~\cite{Sirunyan:2018ygk}.}
\label{fig:HTdistribution}
\end{figure}

\section{SFitter framework}

\begin{table}[b!]
\begin{center}
\begin{tabular}{ll|ll}
\toprule
production & decay & ATLAS & CMS \\ 
\midrule
      & $H \to WW$             & \cite{ATLAS13WW,ATLAS13tthLep}   & \cite{CMS13WW,CMS13tthLep1,CMS13tthLep2} \\
      & $H \to ZZ$             & \cite{ATLAS13ZZ,ATLAS13tthLep}   & \cite{CMS13ZZ, CMS13ZZv2, CMS13tthLep1,CMS13tthLep2} \\
      & $H \to \gamma\gamma$   & \cite{ATLAS13aa}   & \cite{CMS13aa} \\
      & $H \to \tau \tau$      &    \cite{ATLAS13tthLep}                & \cite{CMS13tautau,CMS13tthLep1,CMS13tthLep2} \\
      & $H \to Z\gamma$        & \cite{ATLAS13Za}   & \cite{CMS13Za} \\
\midrule
 WBF  & $H \to \text{inv}$     &                 & \cite{CMS13WBFInv} \\
 WBF  & $H \to \tau \tau$      &               & \cite{CMS13tautau} \\
\midrule
$VH$  & $H \to b\bar{b}$       & \cite{ATLAS13Vhbb} & \cite{CMS13Vhbb} \\
$VH$  & $H \to \tau \tau$      &   & \cite{CMS13Vhtautau} \\
$VH$  & $H \to \text{inv}$     &  \cite{ATLAS13ZhInv}               & \cite{CMS13ZhInv} \\
$VH$  & $H \to b\bar{b}$ ($m_{VH}$) &   & \cite{ATLAS13VhEXO} \\
\midrule
$t\bar{t}H$ & $H \to \gamma\gamma$ & \cite{ATLAS13tthObs}   & \cite{CMS13aa}     \\
$t\bar{t}H$ & $H \to ZZ \to 4 \ell$ & \cite{ATLAS13tthObs}     & \cite{CMS13ZZ, CMS13ZZv2}   \\
$t\bar{t}H$ & $H \to WW,ZZ,\tau\tau$& \cite{ATLAS13tthLep} & \cite{CMS13tthLep1,CMS13tthLep2} \\
$t\bar{t}H$ &$H \to b\bar{b}$      &   \cite{ATLAS13tthbb}     & \cite{CMS13tthbb} \\
\bottomrule
\end{tabular}
\end{center}
\caption{List of Run~II Higgs measurements included in our
  analysis. For the $m_{VH}$ distribution our highest-momentum bin
  with observed events starts at $m_{VH} = 990$~GeV and $1.2$~TeV
  for the $0 \ell$ and $1 \ell$ final states.}
\label{tab:higgs_data}
\end{table}

In \textsc{SFitter} analyses we prefer not to rely on the
pre-processed rate modifiers by ATLAS and CMS whenever
possible. Instead, we extract the signal and background rates from the
experimental publications and apply our own uncertainty
treatment. This includes correlated and uncorrelated systematic
uncertainties as well as a flat likelihood within the allowed band by
theoretical uncertainties. For analyses using multivariate analysis
techniques, where the number of events in each signal region is only
illustrated after simple cuts rather than the full analysis, we
implement the signal strength modifiers but separate for example the
theory uncertainties.  All signal efficiencies and higher-order
effects we extract as the difference between our simulation and the
numbers quoted by ATLAS and CMS. 

For Higgs and di-boson signals we use
\textsc{MadGraph5}~\cite{madgraph} for the event generation,
\textsc{Pythia6}~\cite{pythia} for parton shower and hadronization,
and \textsc{Delphes3}~\cite{delphes} for the detector simulation.
Branching ratios including dimension-6 effects are given by the
extended version of \textsc{Hdecay}~\cite{hdecay}. For new physics
effects in the production process we use the same tool chain as for
the Standard Model, combined with our
\textsc{FeynRules}~\cite{feynrules} implementation of the dimension-6
operators and assume that detector effects as well as higher-order
corrections scale with the SM case in the fiducial volume of the
SM-like measurement. For total rate measurements using the bulk of the
phase space this approximation is obviously justified. For our
kinematic distributions this is less clear, so we have checked that
our approach is approximately correct~\cite{fermionic,sally}.
Corrections to diboson production have been calculated and should 
eventually be included~\cite{dennerNLO}.

As usual for our \textsc{SFitter} analysis we allow for the
modification of the production amplitude through dimension-6
operators including the interference with the SM amplitude and
the squared term in the Wilson coefficient. The latter becomes
relevant whenever the interference with the Standard Model is
suppressed. Given the estimates of Eq.\eqref{eq:rough_higgs} and
Eq.\eqref{eq:rough_ewpo} we simplify our analysis by neglecting
diagrams which are modified by bosonic and fermionic operators at the
same time and interfere with the SM amplitude. In our discussion of
the results we will see that indeed large effects from the fermionic
operators do not appear in this topology. Finally, we neglect
dimension-6 squared contributions of the fermionic operators to the
gauge boson branching ratios, because they will be strongly suppressed
following Eq.\eqref{eq:rough_ewpo} with a typical energy scale $m_V$
in the gauge boson decays. For the same reason we neglect the effects
of the fermionic operators on the decays of gauge bosons coming from
Higgs decays. The hierarchy of scales combined with the well-defined
external energy scale $E \lesssim m_H$ will render them numerically
irrelevant.\medskip

For Higgs and di-boson we start with the set of Run~I measurements
discussed in Refs~\cite{legacy1,legacy2}. We add the Run~II Higgs
measurements shown in Tab.~\ref{tab:higgs_data} and the Run~II
di-boson measurements shown in Tab.~\ref{tab:vv_data}.  Because the
dimension-6 Lagrangian introduces new Lorentz structures and hence
predicts significantly different event kinematics from the Standard
Model, kinematic distributions scaling with energy are especially
powerful. An attractive case is a $m_{VH}$ distribution from an ATLAS
resonance search~\cite{ATLAS13VhEXO}, which we include for the
zero-lepton and one-lepton final states. We re-bin the reported result
such that the most relevant high bins include a statistically
meaningful number of events, giving us measurements exceeding $m_{VH}
= 1$~TeV.\footnote{We would happily thank ATLAS for help with this
  analyis result and we are grateful to the actual authors
  communicating with us. However, our EFT analysis was officially
  considered as no appropriate re-casting of a $VH$ resonance search,
  so there is nothing we can thank ATLAS for.}  The other side of the
kinematics medal is that differential measurements from $H \to 4\ell$ decays can be
safely neglected in a global analysis. The reason is that the momentum
flow through the Higgs decay vertex is cut off by the on-shell
condition, so any measurement in $VH$ or WBF production will surpass
their impact on a global analysis~\cite{info_geo}.

Based on all measurements we first construct a multi-dimensional, full
exclusive likelihood map. As long as we are only interested in small
deviations from the Standard Model, a key assumption to be able to use
an effective field theory approach, we can assume that local SM-like
minima are also the global minima in this likelihood map.  There exist
three standard ways to explore the log-likelihood distribution around
the minimum: first, we can use a naive, \textsc{Minuit}-like approach,
approximating the functional form around the minimum by a quadratic
function. This assumption is not appropriate once we allow for
non-Gaussian errors, for example a flat shape covering the theoretical
error bar. Second, we can construct a Markov chain over the parameter
space. Here the problem is that different directions in the space of
Wilson coefficients behave differently, which makes it hard to define
a universal and efficient proposal function. Nevertheless, we check
our results against such a Markov chain analysis and usually find
encouraging agreement.  For our numerical analysis we define 10.000
toy measurements, modeling the Poisson, Gaussian or flat input
distributions. For each toy experiment we determine the best-fitting
point in the space of Wilson coefficients, combine these values to a
histogram, effectively profile over the remaining parameters, and
determine the 68\% and 95\% ranges around the SM-like central
value. For the error bands we require the log-likelihood values at the
lower and upper ends to be identical. 

\begin{figure}[t]
\includegraphics[width=0.45\textwidth]{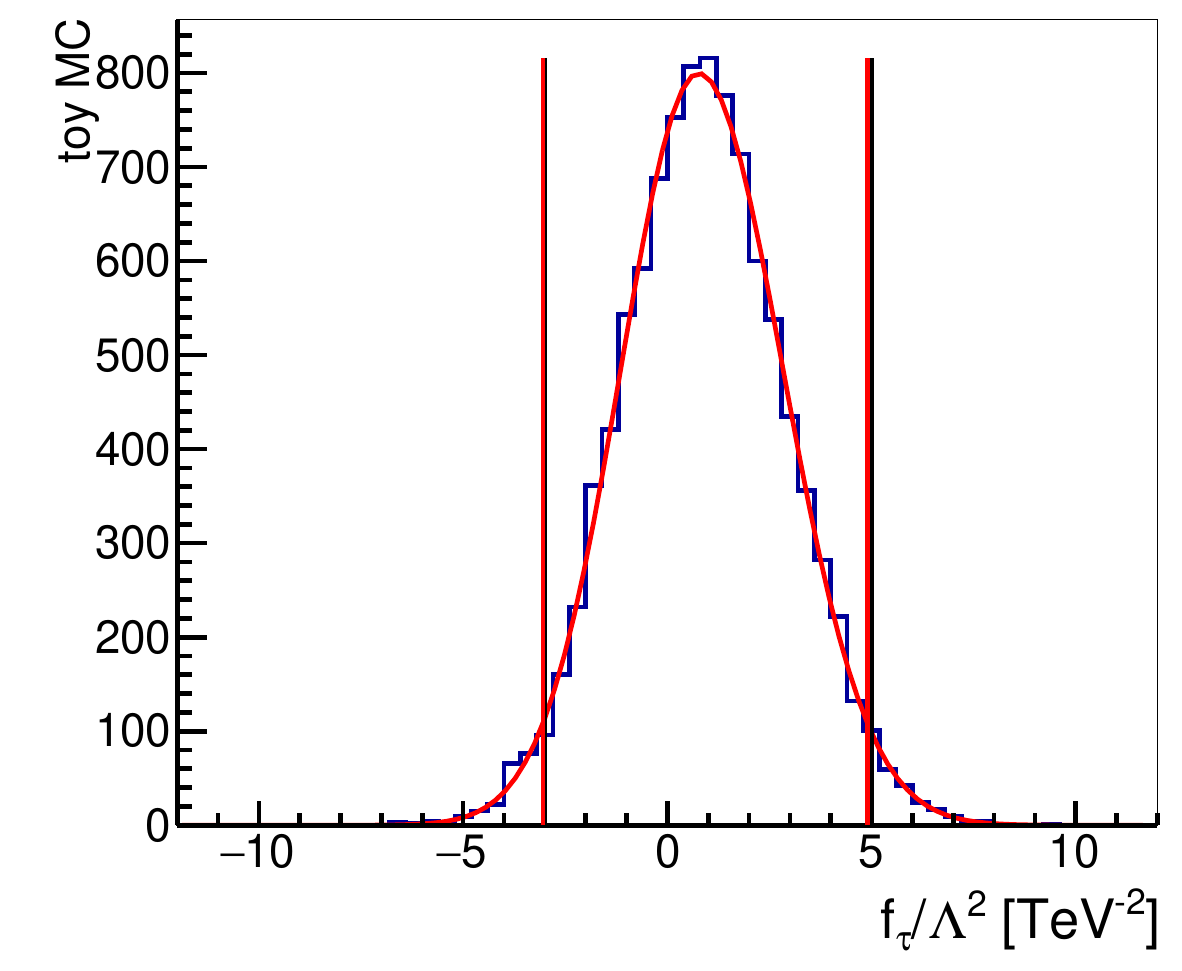}
\hspace*{0.05\textwidth}
\includegraphics[width=0.45\textwidth]{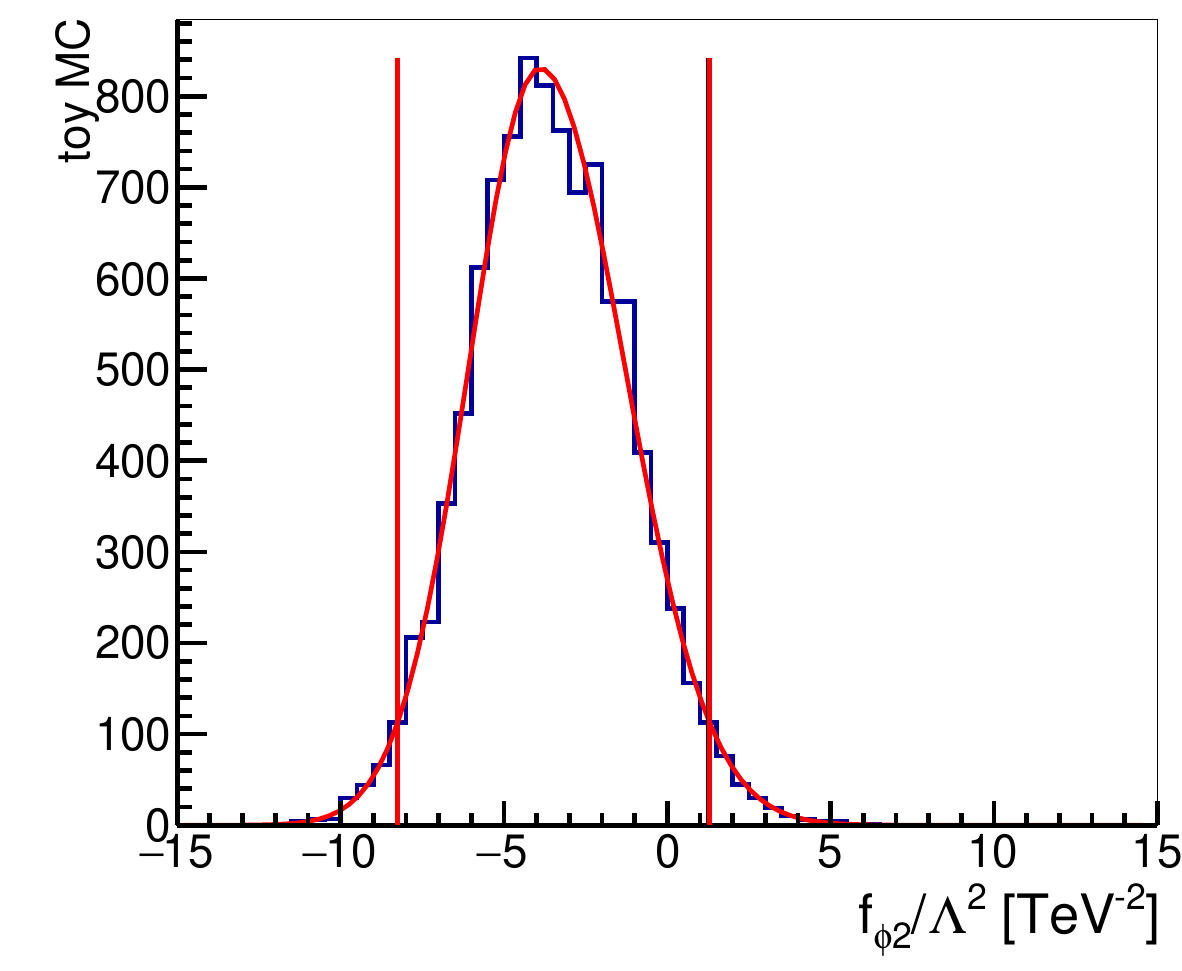}\\
\includegraphics[width=0.45\textwidth]{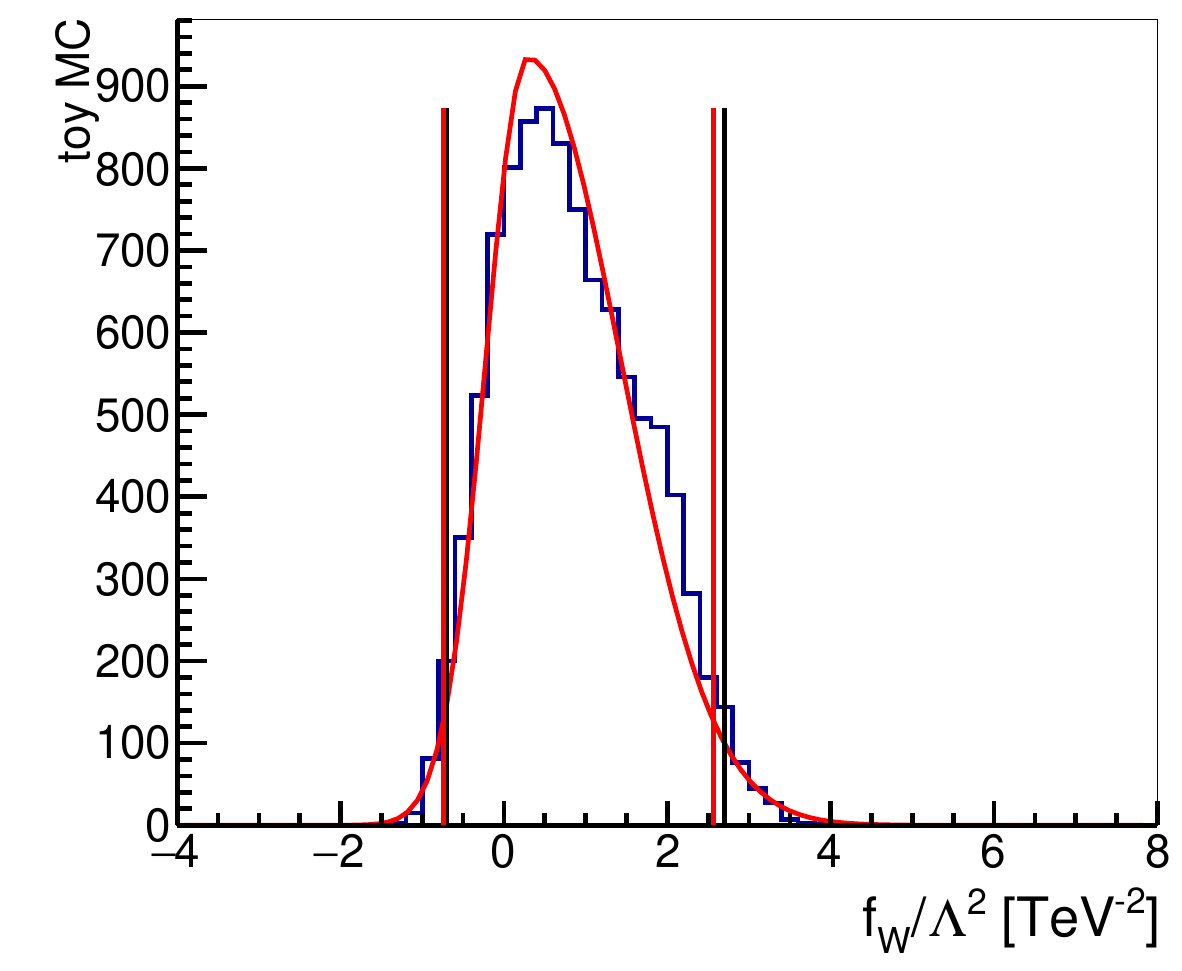}
\hspace*{0.05\textwidth}
\includegraphics[width=0.45\textwidth]{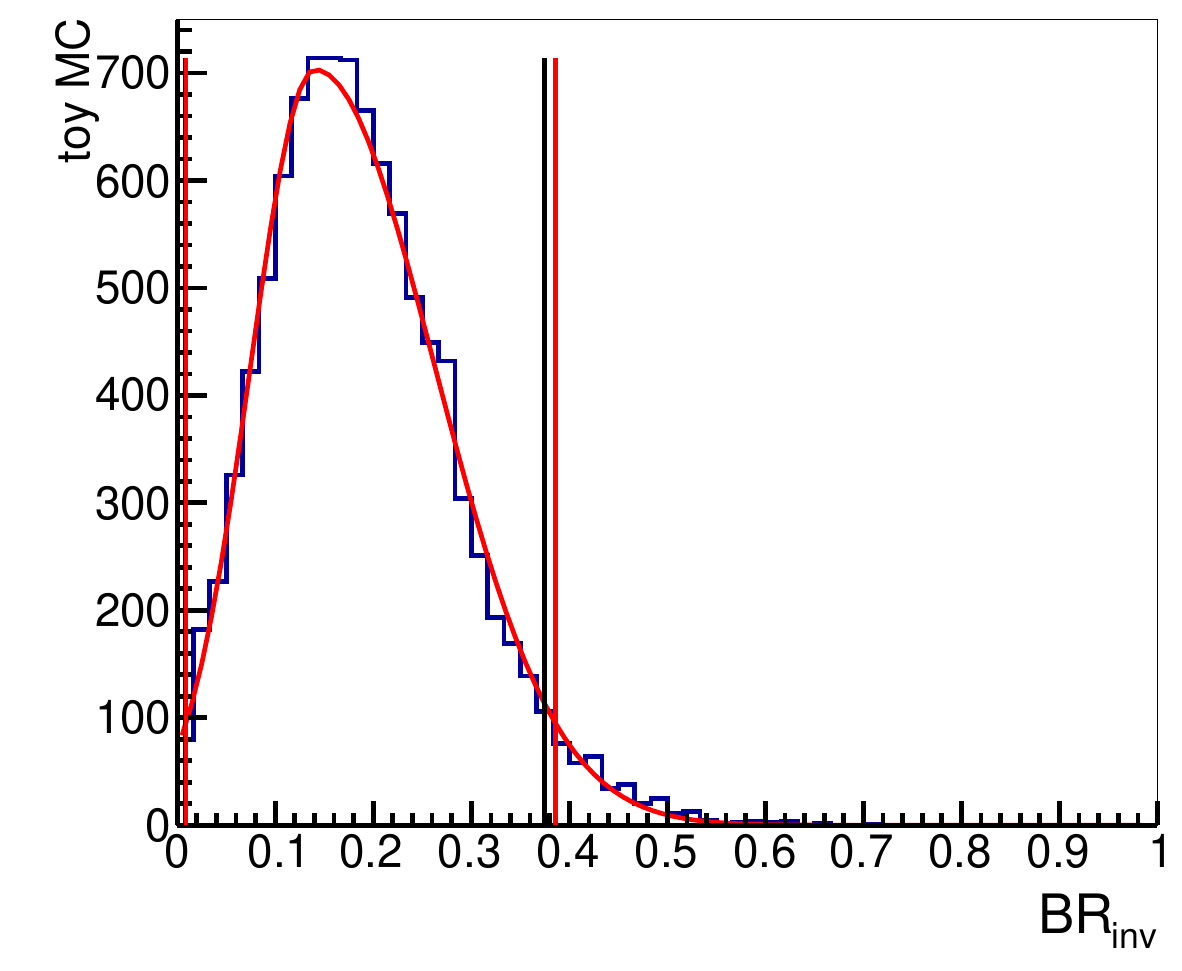}
\caption{Distributions of the toy experiments for the operators
  $\ope_\tau$, $\ope_{\phi 2}$ and $\ope_W$
  as well as the invisible Higgs branching ratio, based on the full LHC
  data set. The lines show the 95\% CL limits from the histogram
  (black) and the double-Gaussian fit (red).}
\label{fig:toys}
\end{figure}

Because our approch gives us full control over the log-likelihood
distribution we can compare these limits with a dual Gaussian fit to
the log-likelihood in one dimension.  We find good agreement for all
Wilson coefficients, even though Fig.~\ref{fig:toys} shows that for
example the profile likelihood for $f_W$ does not have a symmetric
Gaussian shape. Obviously, the shape for the invisible Higgs width is
distorted, because it does not allow for negative branching ratios.
While we quote the error bars for the non-Gaussian analysis we quote
the results from the Gaussian fit whenever we give a best-fit point
for a Wilson coefficient.  For additional details on the
\textsc{SFitter} framework we refer to
Refs.~\cite{sfitter_orig,sfitter_delta}.\medskip

\begin{table}[b!]
\begin{center} 
\begin{tabular}{ll|lccr}
\toprule%
&channel & distribution & \#bins & max [GeV] & \\
\midrule
\multirow{4}{*}{8~TeV}
&$WW\rightarrow \ell^+\ell^{\prime -}+\met \; (0j)$    & leading $p_{T,\ell}$ & 4 & 350 & $20.3~\ifb$~\cite{atlas8ww}  \\
&$WW\rightarrow \ell^+\ell^{(\prime) -}+\met \; (0j)$  & $m_{\ell\ell^{(\prime)}}$ & 7 & 575 & $19.4~\ifb$~\cite{cms8ww}  \\
&$WZ\rightarrow \ell^+\ell^{-}\ell^{(\prime)\pm}$       & $m_T^{WZ}$           & 6 & 450 & $20.3~\ifb$~\cite{atlas8wz}  \\
&$WZ\rightarrow \ell^+\ell^{-}\ell^{(\prime)\pm}+\met $ & $p_T^{Z \to \ell\ell}$& 8 & 350 & $19.6~\ifb$~\cite{cms78wz}  \\
\midrule
13~TeV
&$WZ\rightarrow \ell^+\ell^{-}\ell^{(\prime)\pm}$       & $m_T^{WZ}$           & 7 & 675 & $36.1~\ifb$~\cite{ATLAS13WZ}  \\
\bottomrule
\end{tabular} 
\end{center}
\caption{List of Run~I and Run~II di-boson measurements included in
  our analysis. The maximum value in GeV indicates the lower end of
  the highest-momentum bin we consider.}
\label{tab:vv_data}
\end{table}

One caveat applies to all analyses based on effective Lagrangians: we
consider the dimension-6 Lagrangian of Eq.\eqref{eq:ourlag1} and
Eq.\eqref{eq:ourlag2} the appropriate desciption of the physics
effects beyond the Standard Model. Note that this statement by no
means implies that for example the dimension-6-squared contributions
have to be smaller than those from the dimension-6 interference with
the Standard Model~\cite{interference}. There exist many physics
reasons why this could be a valid physics effect, and the discrepancy
between the generic LHC reach given by Eq.\eqref{eq:rough_higgs} and
the generic reach of electroweak precision data in
Eq.\eqref{eq:rough_ewpo} will be discussed as an example for such
effects in the next section.  Instead, we simply need to ensure that
no particle of the UV-completions which we approximate with our
effective Lagrangian contributes as a propagating degree of freedom on
its mass shell~\cite{validity}. To this end, computing the effects of
dimension-8 operators can give useful hints about the validity of the
dimension-6 truncation~\cite{dim8}, but it does not have to.

Finally, in the spirit of the effective field theory we only consider
SM-like scenarions, which means that we neglect all secondary
solutions for example with switched signs of Yukawa couplings.
Assuming weakly interacting new physics such effects require scales
$\Lambda \sim m_H$, so we expect these models to be best tested in
direct LHC searches rather than a global analysis.  In any case, the
observation of a sign switch for example in a Yukawa coupling as part
of a global analysis would signal a breakdown of the renormalizable
Standard Model and its symmetry structure and would prompt us to
modify our SMEFT hypothesis.  Of course, when it comes to searching
for effects in kinematic distributions, these two search strategies
are closely related, for example when we directly search for mass
peaks in the same distributions that we indirectly test for shoulders
(as an early sign of a mass peak appearing in data)~\cite{validity}.

\section{Global analysis}
\label{sec:higgs}

\begin{figure}[t]
\centering
\includegraphics[width=0.7\textwidth]{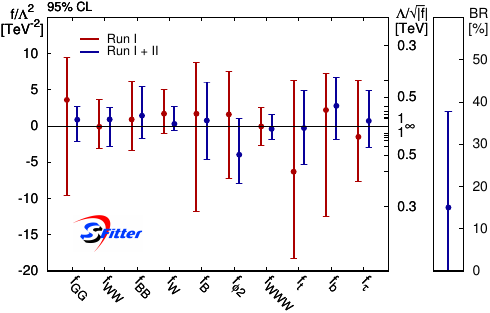} 
\caption{Allowed 95\% CL ranges for individual Wilson coefficients
  $f_x/\Lambda^2$ from a one-dimensional profile likelihood. We show
  results from Run~I (red) and using the additional Run~II
  measurements (blue). We neglect all operators contributing to
  electroweak precision observables at tree level.}
\label{fig:global_noewpo}
\end{figure}

Before we attempt a proper global analysis of the Higgs and
electroweak gauge sector we can ask what the impact of the additional
13~TeV data given in Tabs.~\ref{tab:higgs_data} and~\ref{tab:vv_data}
is. Aside from a generic improvement in many of the standard
measurements, we expect a significant impact from the new $t\bar{t}H$
measurements, the significant observation of fermionic Higgs decays,
and from the re-casted $m_{VH}$ distribution to very large
energies. In Fig.~\ref{fig:global_noewpo} we indeed see that the
limits on $f_t$, $f_b$ and $f_\tau$ have improved by more than a
factor of two. Obviously, the top Yukawa measurement directly affects
the Higgs coupling to gluons, $\ope_{GG}$, because it can only be
extracted after we subtract the measured top loop contribution.
Because $\ope_{\phi 2}$ leads to a Higgs wave function renormalization
and $\ope_b$ modifies the total Higgs width, they are strongly
correlated in the global analysis. After Run~II they are not only well
determined, both of them also show symmetric Gaussian log-likelihood
distributions. We also see a very significant improvement in the limit
on $f_W$ and $f_B$, which is driven by associated $VH$ production.
However, from Fig.~\ref{fig:toys} we know that the error bar on $f_W$
is by no means symmetric and Gaussian due to 
the relative size of the linear and quadratic terms of the EFT, the 
parametrization of the theory prediction and further effects. 
The operators showing the
least improvement compared to Run~I are $\ope_{WW}$ and $\ope_{BB}$,
reflecting the lack of high-impact kinematic WBF measurements in the
Run~II data set.  Moreover, $\ope_{WWW}$ only affects the gauge
sector, and in Tab.~\ref{tab:vv_data} we see that the analysis is
still dominated by a broad set of extremely successful kinematic
measurements at Run~I in view of a global gauge analysis.

Finally, our global limit on the Higgs branching ratio to invisible
particles is
\begin{align}
\text{BR}_\text{inv} < 38\%
\qquad \text{at $95\%$~CL,}
\end{align}
with a best-fit point of $\text{BR}_\text{inv} = 14\%$. This is
significantly weaker than the limits quoted for example by
CMS~\cite{CMS13WBFInv}, because our global analysis does not assume
the underlying Higgs production rates to be SM-like. Indeed, we
observe a strong correlation of the invisible branching ratio with
$\ope_{\phi 2}$ and its universal Higgs wave function
renormalization. If rather than profiling over it we fix $f_{\phi 2} =
0$, our limit becomes $\text{BR}_\text{inv} < 26\%$ in agreement with
the experimental results.  Altogether, we find that Run~II
systematically probes energy scales $\Lambda/\sqrt{f}$ between 400~GeV
and 800~GeV through Higgs measurement.\medskip

\begin{figure}[t]
\centering
\includegraphics[width=.65\textwidth]{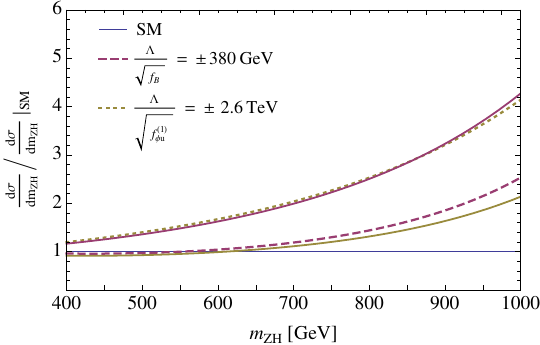}
\caption{Invariant mass distribution $m_{ZH}$ normalized to the
  Standard Model.  The dashed lines correspond to
  $\Lambda/\sqrt{|f_B|} = +380$~GeV and $\Lambda/\sqrt{|f^{(1)}_{\phi
      u}|} = +2.6$~TeV, while the solid lines correspond to the negative 
      values of the Wilson coefficients with the same magnitude. }
\label{fig:mzh_dist}
\end{figure}

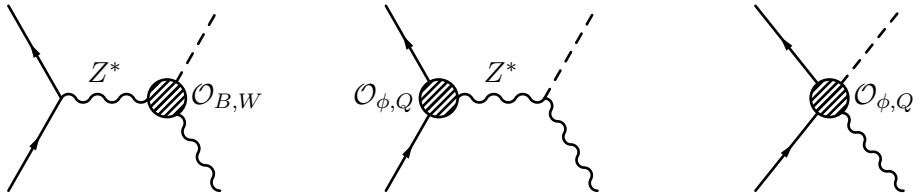
\begin{figure}[b!]
\centering
\begin{fmfgraph*}(100,70)
\fmfset{arrow_len}{2mm}
\fmfleft{i1,i2}
\fmfright{o1,o2}
\fmf{fermion}{i1,v1}
\fmf{fermion}{v1,i2}
\fmf{photon,tension=1,label=$Z^*$,lab.side=left,width=1}{v1,v2}
\fmf{photon,width=1}{v2,o1}
\fmf{dashes}{v2,o2}
\fmflabel{$\;\ope_{B,W}$}{v2}
\fmfv{decor.shape=circle,decor.filled=shaded}{v2}
\end{fmfgraph*}
\hspace*{0.08\textwidth}
\begin{fmfgraph*}(100,70)
\fmfset{arrow_len}{2mm}
\fmfleft{i1,i2}
\fmfright{o1,o2}
\fmf{fermion}{i1,v1}
\fmf{fermion}{v1,i2}
\fmf{photon,tension=1,label=$Z^*$,lab.side=left,width=1}{v1,v2}
\fmf{photon,width=1}{v2,o1}
\fmf{dashes}{v2,o2}
\fmflabel{$\ope_{\phi Q} \;$}{v1}
\fmfv{decoration.shape=circle,decoration.filled=shaded}{v1}
\end{fmfgraph*}
\hspace*{0.08\textwidth}
\begin{fmfgraph*}(70,70)
\fmfset{arrow_len}{2mm}
\fmfleft{i1,i2}
\fmfright{o1,o2}
\fmf{fermion}{i1,v1}
\fmf{fermion}{v1,i2}
\fmf{photon,width=1}{v1,o1}
\fmf{dashes}{v1,o2}
\fmflabel{$\; \ope_{\phi Q}$}{v1}
\fmfv{decoration.shape=circle,decoration.filled=shaded}{v1}
\end{fmfgraph*}
\caption{Dimension-6 contribution to $ZH$ production. We show sample
  diagrams for the usual bosonic corrections, the small fermionic
  corrections from a 3-point vertex, and the large fermionic
  corrections from a 4-point interaction.}
\label{fig:feyn_vh}
\end{figure}

\begin{figure}[t]
\centering
\includegraphics[width=0.35\textwidth]{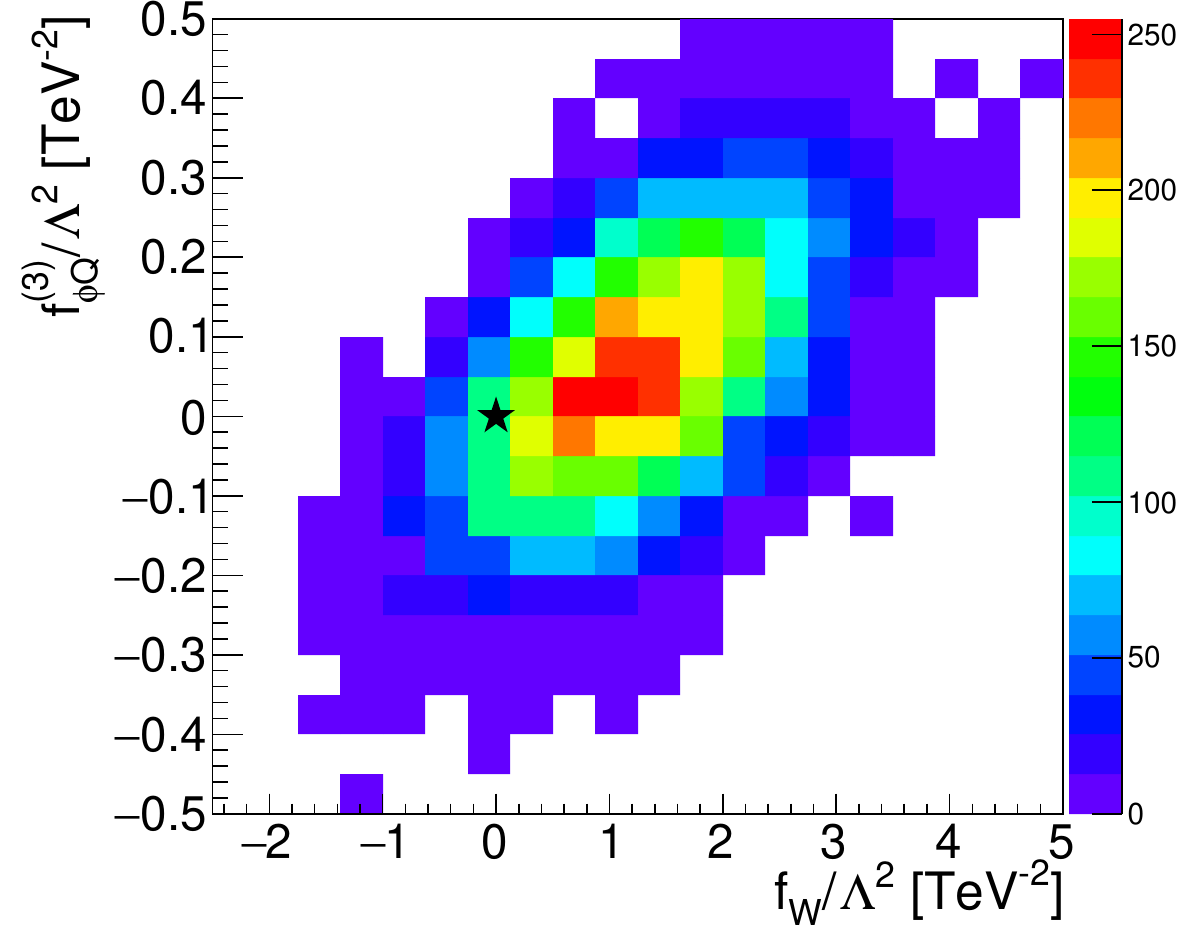} 
\hspace*{0.05\textwidth}
\includegraphics[width=0.35\textwidth]{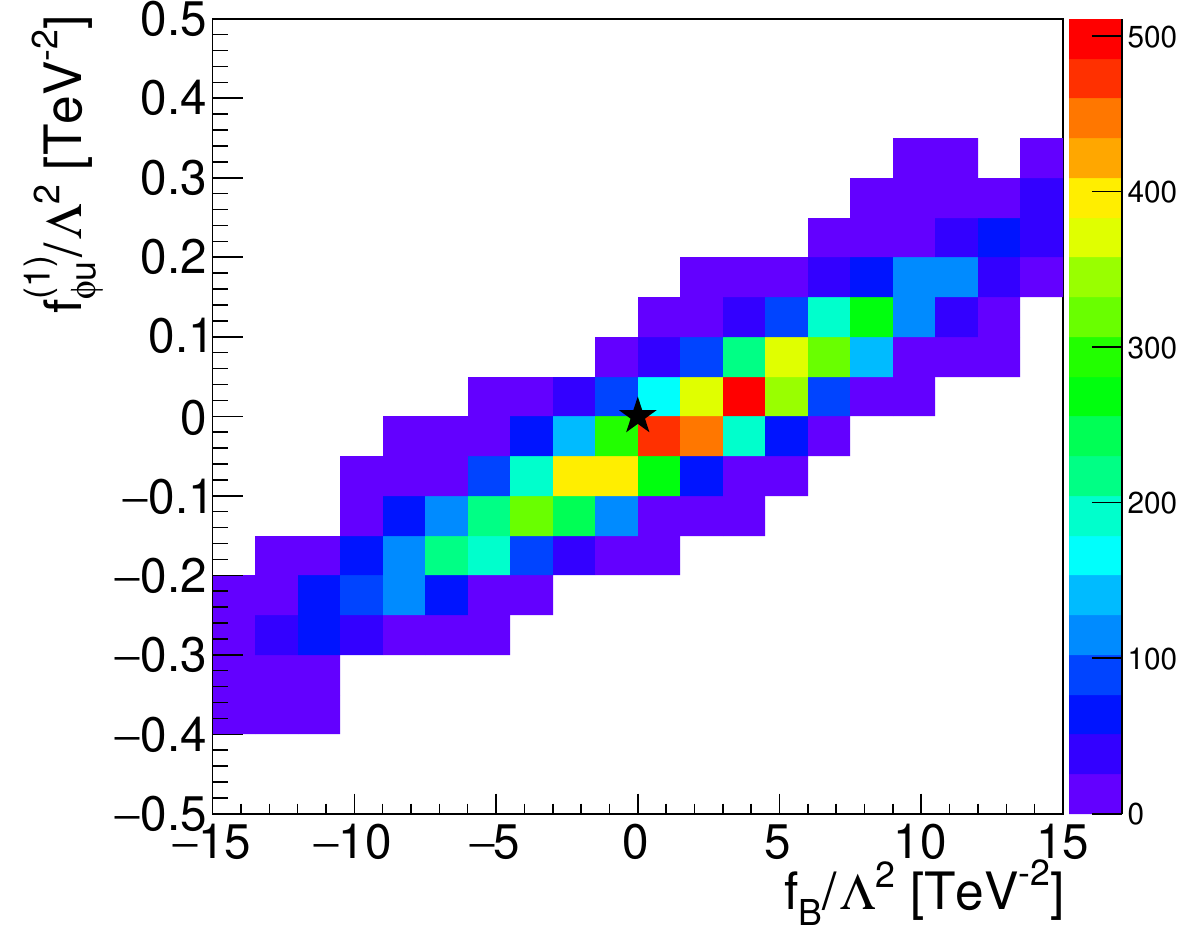} \\
\includegraphics[width=0.35\textwidth]{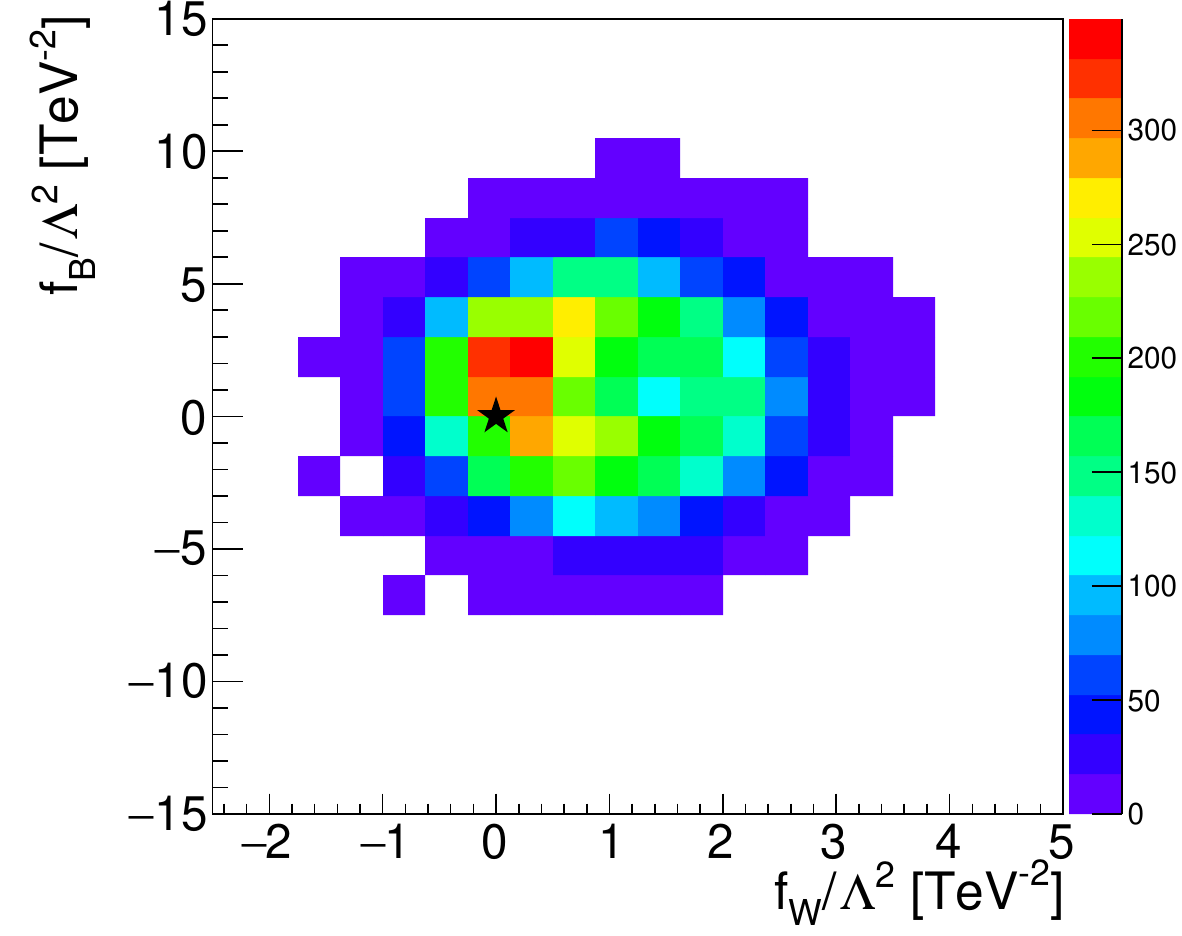} 
\hspace*{0.05\textwidth}
\includegraphics[width=0.35\textwidth]{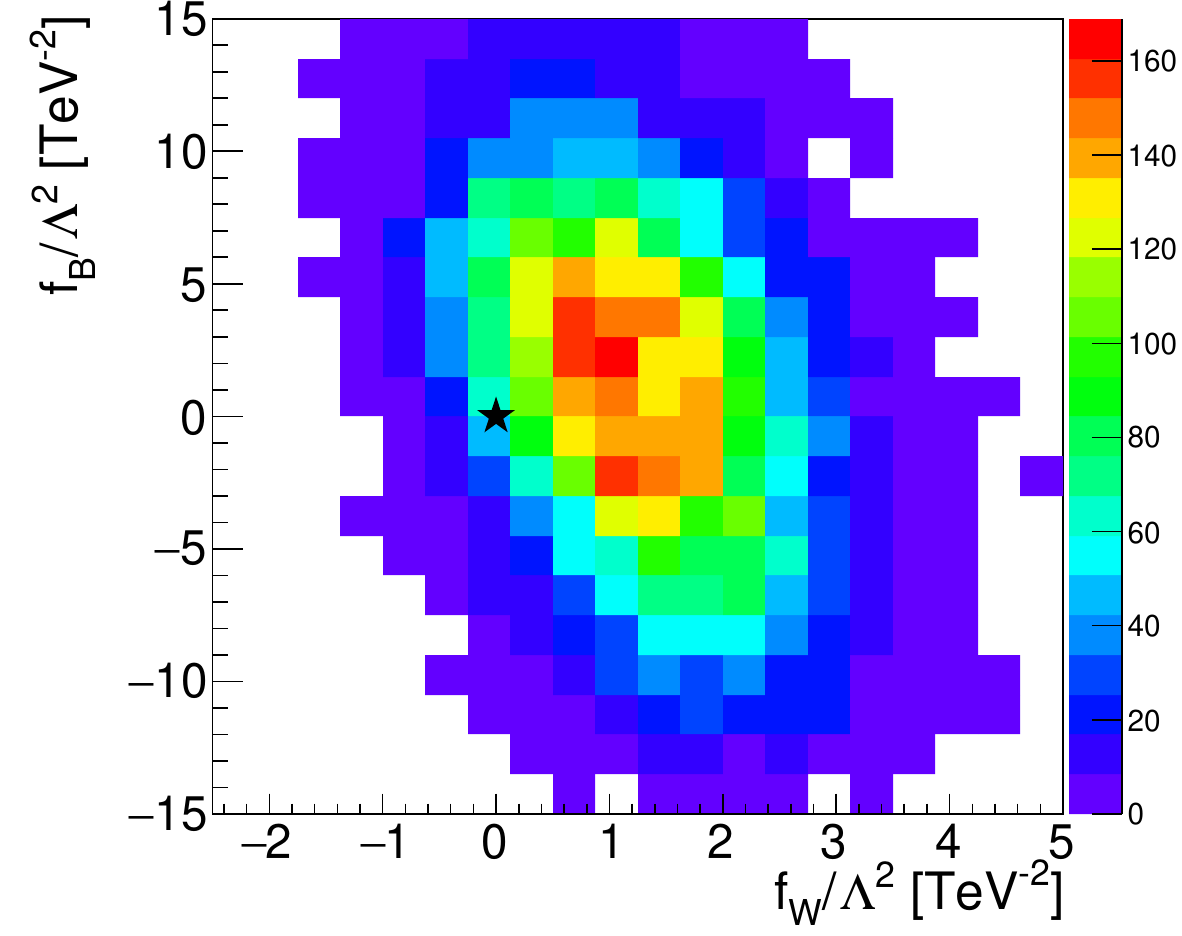} 
\caption{Correlations between the fermionic and bosonic operators (top
  row), and between the usual bosonic operators (bottom row). For the
  latter we show the purely LHC results (left) and the results after
  including the additional fermionic operators.}
 \label{fig:corr_phiq3_w}
\end{figure}

The large improvement of the limits on $\ope_B$ at Run~II forces us to
consider the interplay with the fermionic operators from
Eq.\eqref{eq:ourlag2} and their limits from electroweak precision
data, Eq.\eqref{eq:rough_ewpo}.  From a scale separation point of view
it is seems counter-intuitive that $\ope_{\phi u}^{(1)}$ or
$\ope_{\phi Q}^{(3)}$, for which $\Lambda/\sqrt{f}$ is constrained
around one order of magnitude more strongly than for $\ope_W$ and much
more strongly for all other operators shown in
Fig.~\ref{fig:global_noewpo}, should have any effect on the LHC
analysis~\cite{fermionic}.  In Fig.~\ref{fig:mzh_dist} we see how the
fermionic and bosonic operators affect for example $ZH$
production. The key observation is that the fermionic operator
contributes via the 3-point $qqZ$ and the 4-point $qqHZ$
vertices, whereas the bosonic operators require the same $s$-channel
$Z$-propagator we see in the Standard Model. We show the corresponding
Feynman diagrams in Fig.~\ref{fig:feyn_vh}. From the structure of the
dimension-6 operator we can infer the scalings
\begin{align}
\frac{g f_{\phi Q} \, v^2}{\Lambda^2} 
\quad (qqZ) 
\qqquad \text{versus} \qqquad 
\frac{g f_{\phi Q} \, v}{\Lambda^2} 
\quad (qqZH) \qqquad
\label{eq:energy_scaling}
\end{align}
The $m_{ZH}$ distribution shown in Fig.~\ref{fig:mzh_dist} is one of
our most powerful observables. We have confirmed that for the
fermionic operator it is  entirely dominated by the 4-point
interaction, even though the 3-point interaction does interfere with
the Standard Model. This is due to the suppression of the amplitudes with 
propagating $Z$s due to the off-shell $Z$ which leads to a suppression 
going as $\sim1/(m_{ZH}^2-M_Z^2)<1$ as well as the energy scaling in
Eq.\eqref{eq:energy_scaling} as well as which will eventually also lead to
unitarity violation~\cite{unitarity}.

It is interesting to see how two operators with an apparently very
different new physics scale contribute to the $m_{ZH}$ distribution at
around the same rate. This can be understood by the definitions of the
operators which include a factor of the gauge coupling for each 
field strength tensor. While the 4-point contribution from $\ope_{\phi u}^{(1)}$
lacks a second power of the gauge coupling $g'$ the definition of $\ope_B$
adds two powers of the gauge coupling to the 3-point vertex. Over most of the
parameter range shown in Fig.~\ref{fig:mzh_dist} the
dimension-6-squared contribution dominates, giving us a mis-match of
four powers of the coupling just from the definitions of the Wilson
coefficients.\medskip

We confirm these findings in Fig.~\ref{fig:corr_phiq3_w}, where we
show the resulting correlations in our global analysis, once we
include the full Lagrangian of Eqs.\eqref{eq:ourlag1}
and~\eqref{eq:ourlag2}. We see a clear correlation between $f_B$ and
$f_{\phi u}^{(1)}$ from $ZH$ production, as well as between $f_W$ and
$f_{\phi Q}^{(3)}$ from $WH$ production.  This correlation relates
very different values of the new physics scales for the fermionic and
bosonic operators.  In the lower panels we see how this weakens the
limits on the bosonic operators $f_B$ and $f_W$ after profiling over
the fermionic Wilson coefficients, and how it re-induces a correlation
between them.\medskip

\begin{figure}[t]
\includegraphics[width=0.99\textwidth]{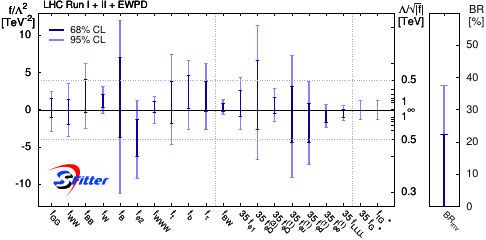} 
\caption{Allowed 68\% and 95\% CL ranges for individual Wilson
  coefficients $f_x/\Lambda^2$ from a one-dimensional profile
  likelihood. All results include the Run~II measurements combined
  with electroweak precision data. We quote the best results for
  $\ope_G$~\cite{krauss} and $\ope_{tG}$~\cite{topfitter} from
  non-Higgs analyses.}
\label{fig:global_all}
\end{figure}

All of this discussion clearly defines a new challenge for global
Higgs analyses once we reach Run~II levels of precision: we need to
include the additional operators shown in
Eq.\eqref{eq:ourlag2}~\cite{fermionic,runII_eng,runII_concha}. As
argued above, this is at least in part due to a relative enhancement
of the fermionic Higgs-gauge operators through their 4-point interactions. We
show the result of our global analyses in Fig.~\ref{fig:global_all},
both at the 68\% and 95\% confidence levels.  As LHC observables we
consider the same measurements as Fig.~\ref{fig:global_noewpo}, but
now combined with electroweak precision observables and including an
extended set of operators. While the triple-gluon operator $\ope_G$
and the chromo-magnetic operator $\ope_{tG}$ appear in a global Higgs
analysis, we have shown in Sec.~\ref{sec:top_sector} that their best
limits come from dedicated studies and after considering these limits
their effects on the Higgs observables will not be visible. We
therefore include them in the SMEFT-like result shown in
Fig.~\ref{fig:global_all}, but quote the constraints from non-Higgs
analyses.

First, we see that the 68\% and 95\% confidence limits scale like we
would expect from Gaussian uncertainties.  Directly comparing the
results for the bosonic operators without and with the fermionic
operators we see that as expected from Fig.~\ref{fig:corr_phiq3_w} the
results on $f_B$ are roughly a factor of two weaker once we profile
over the fermionic Wilson coefficients. We also see weaker limits on
$f_W$ and $f_{\phi 2}$, which propagate through the entire effective
Lagrangian describing the global analysis. 

The constraints from our global analysis on the fermionic Higgs-gauge
operators are typically a factor 10 to 100 stronger than for the
bosonic operators. With $f_{\phi Q}^{(3)}$ and $f_{\phi d}^{(1)}$ the
global fit also constrains operators which are relatively poorly
probed by electroweak precision observables alone. These limits are in
the range of $\Lambda/\sqrt{f} \approx 3$~TeV at 68\%~CL, indicating
that LHC observables can also be especially sensitive to these
operators. Again, for these results it is crucial that our global
Higgs analysis covers Higgs observables and di-boson observables at
the LHC, combined with electroweak precision data.

\section{Summary}
\label{sec:conclu}

We have presented a global analysis of the LHC Run~I and Run~II
measurements related to Higgs and di-boson measurements in the
framework of an effective Lagrangian to dimension six.  The
increasingly strong constraints from Run~II and especially the
developing LHC sensitivity to anomalous gauge boson couplings to
quarks require a combination of the LHC analysis with electroweak
precision data. In our global Higgs and electroweak analysis we
include 18 bosonic and fermionic dimension-6 operators. For two more
operators we quote limits from other analyses, after
confirming that they are more constraining than our Higgs
analysis. Finally, we include invisible Higgs decays through their
branching ratio. This set of operators defines a significant step
towards a global SMEFT analysis in the LHC era and towards a global
precision analysis of LHC data.\medskip

In the \textsc{SFitter} framework we directly analyze ATLAS and CMS
measurements rather than pre-defined pseudo-observables, include
correlations for systematic and theoretical uncertainties, and exploit
kinematic distributions to large momentum transfer.  For LHC data
alone we find that all limits from Run~I are consistently improved by
Run~II, especially in the Yukawa sector and from the kinematic
measurements of $VH$ production. At 95\% CL the typical Run~II limits
range around $\Lambda/\sqrt{f} = 400~...~800$~GeV. Through new 4-point
vertices fermionic Higgs-gauge operators have an anomalously large
effect on associated Higgs production. This induces strong
correlations between fermionic operators and $f_{B,W}$, in spite of
stringent constraints from electroweak precision data. Profiling over
the fermionic Wilson coefficients weakens the limits on $f_B$ by a
factor two. At the same time, LHC observables allow us to constrain
fermionic operators like $f_{\phi d}^{(1)}$ far beyond the reach of
electroweak precision data, indicating that the interaction between
the two sectors of our global fit is mutual. For several bosonic
operators our analysis probes $\Lambda/\sqrt{f}$ values up to the TeV
range, while the fermionic Higgs-gauge operators are consistently
constrained to $5~...~10$~TeV.

\bigskip
\begin{center} \textbf{Acknowledgments} \end{center}

We are grateful to Dirk Zerwas and Michael Rauch for their continuous
support of \textsc{SFitter} and for many discussions related to LHC
measurements.  We would like to warmly thank Tatjana Lenz and Ruth
Jacobs for their help with the $m_{VH}$ distribution.  We would like
to thank the DAAD Australia exchange program for funding our project
\textsl{Precision Higgs Physics at the LHC} (57390316).  AB is funded
through the Graduiertenkolleg \textsl{Particle physics
  beyond the Standard Model} (GRK 1940). TC acknowledges generous
support from the Villum Fonden and partial support by the Danish
National Research Foundation (DNRF91) through the Discovery centre.
The authors acknowledge support by the state of Baden-W\"urttemberg
through bwHPC and the German Research Foundation (DFG) through grant
no INST 39/963-1 FUGG (bwForCluster NEMO).

\end{fmffile}


\begin{thebibliography}{99}

\bibitem{higgs}
  P.~W.~Higgs,
  Phys.\ Rev.\ Lett.\  {\bf 13}, 508 (1964)
  \doi{10.1103/PhysRevLett.13.508};
  P.~W.~Higgs,
  Phys.\ Lett.\  {\bf 12}, 132 (1964)
  \doi{10.1016/0031-9163(64)91136-9};
  F.~Englert and R.~Brout,
  Phys.\ Rev.\ Lett.\  {\bf 13}, 321 (1964)
  \doi{10.1103/PhysRevLett.13.321}.

\bibitem{review} 
  S.~Dawson, C.~Englert and T.~Plehn,
  \arxiv{1808.01324} [hep-ph];
  C.~Englert, A.~Freitas, M.~M.~M\"uhlleitner, T.~Plehn, M.~Rauch, M.~Spira and K.~Walz,
  J.\ Phys.\ G {\bf 41}, 113001 (2014)
  \doi{10.1088/0954-3899/41/11/113001}
  [\arxiv{1403.7191} [hep-ph]].

\bibitem{eft1} 
  S.~Weinberg,
  Physica A {\bf 96}, no. 1-2, 327 (1979)
  \doi{10.1016/0378-4371(79)90223-1};
  see also 
  H.~Georgi,
  Menlo Park, USA: Benjamin/cummings (1984)
  J.~F.~Donoghue, E.~Golowich and B.~R.~Holstein,
  Camb.\ Monogr.\ Part.\ Phys.\ Nucl.\ Phys.\ Cosmol.\  {\bf 2}, 1 (1992)
  [Camb.\ Monogr.\ Part.\ Phys.\ Nucl.\ Phys.\ Cosmol.\  {\bf 35} (2014)].
  \doi{10.1017/CBO9780511524370}.

 \bibitem{eft2}
  C.~N.~Leung, S.~T.~Love and S.~Rao,
  Z.\ Phys.\ C {\bf 31}, 433 (1986)
  \doi{10.1007/BF01588041};
  W.~Buchmuller and D.~Wyler,
  Nucl.\ Phys.\ B {\bf 268}, 621 (1986)
  \doi{10.1016/0550-3213(86)90262-2};
  M.~C.~Gonzalez-Garcia,
  Int.\ J.\ Mod.\ Phys.\ A {\bf 14}, 3121 (1999)
  \doi{10.1142/S0217751X99001494}
  [\arxiv{hep-ph/9902321}];
  B.~Grzadkowski, M.~Iskrzynski, M.~Misiak and J.~Rosiek,
  JHEP {\bf 1010}, 085 (2010)
  \doi{10.1007/JHEP10(2010)085}
  [\arxiv{1008.4884} [hep-ph]];
  G.~Passarino,
  Nucl.\ Phys.\ B {\bf 868}, 416 (2013)
  \doi{10.1016/j.nuclphysb.2012.11.018}
  [\arxiv{1209.5538} [hep-ph]].

 \bibitem{kaoru} 
  K.~Hagiwara, R.~Szalapski and D.~Zeppenfeld,
  Phys.\ Lett.\ B {\bf 318}, 155 (1993)
  \doi{10.1016/0370-2693(93)91799-S}
  [\arxiv{hep-ph/9308347}].

\bibitem{kilian}
  For a pedagogical introduction see \eg
  W.~Kilian,
  Springer Tracts Mod.\ Phys.\  {\bf 198}, 1 (2003)
  \doi{10.1007/b97367}.

\bibitem{trott} 
  I.~Brivio and M.~Trott,
  \arxiv{1706.08945} [hep-ph].

\bibitem{linear}
  A.~De Rujula, M.~B.~Gavela, P.~Hernandez and E.~Masso,
  Nucl.\ Phys.\ B {\bf 384}, 3 (1992)
  \doi{10.1016/0550-3213(92)90460-S};
  K.~Hagiwara, S.~Ishihara, R.~Szalapski and D.~Zeppenfeld,
  Phys.\ Rev.\ D {\bf 48}, 2182 (1993)
  \doi{10.1103/PhysRevD.48.2182};
  K.~Hagiwara, S.~Matsumoto and R.~Szalapski,
  Phys.\ Lett.\ B {\bf 357}, 411 (1995)
  \doi{10.1016/0370-2693(95)00925-B}
  [\arxiv{hep-ph/9505322}];
  K.~Hagiwara, T.~Hatsukano, S.~Ishihara and R.~Szalapski,
  Nucl.\ Phys.\ B {\bf 496}, 66 (1997)
  \doi{10.1016/S0550-3213(97)00208-3}
  [\arxiv{hep-ph/9612268}].

\bibitem{legacy1}
  T.~Corbett, O.~J.~P.~Eboli, D.~Goncalves, J.~Gonzalez-Fraile, T.~Plehn and M.~Rauch,
  JHEP {\bf 1508}, 156 (2015)
  \doi{10.1007/JHEP08(2015)156}
  [\arxiv{1505.05516} [hep-ph]];
  T.~Corbett, O.~J.~P.~Eboli, D.~Goncalves, J.~Gonzalez-Fraile, T.~Plehn and M.~Rauch,
  \arxiv{1511.08188} [hep-ph].

\bibitem{legacy2}
  A.~Butter, O.~J.~P.~Eboli, J.~Gonzalez-Fraile, M.~C.~Gonzalez-Garcia, T.~Plehn and M.~Rauch,
  JHEP {\bf 1607}, 152 (2016)
  \doi{10.1007/JHEP07(2016)152}
  [\arxiv{1604.03105} [hep-ph]].

\bibitem{barca}
  T.~Corbett, O.~J.~P.~Eboli, J.~Gonzalez-Fraile and M.~C.~Gonzalez-Garcia,
  Phys.\ Rev.\ D {\bf 86}, 075013 (2012)
  \doi{10.1103/PhysRevD.86.075013}
  [\arxiv{1207.1344} [hep-ph]];
  T.~Corbett, O.~J.~P.~Eboli, J.~Gonzalez-Fraile and M.~C.~Gonzalez-Garcia,
  Phys.\ Rev.\ D {\bf 87}, 015022 (2013)
  \doi{10.1103/PhysRevD.87.015022}
  [\arxiv{1211.4580} [hep-ph]].

\bibitem{runI_th}
  G.~Belanger, B.~Dumont, U.~Ellwanger, J.~F.~Gunion and S.~Kraml,
  Phys.\ Rev.\ D {\bf 88}, 075008 (2013)
  \doi{10.1103/PhysRevD.88.075008}
  [\arxiv{1306.2941} [hep-ph]];
  P.~P.~Giardino, K.~Kannike, I.~Masina, M.~Raidal and A.~Strumia,
  JHEP {\bf 1405}, 046 (2014)
  \doi{10.1007/JHEP05(2014)046}
  [\arxiv{1303.3570} [hep-ph]];
  P.~Bechtle, S.~Heinemeyer, O.~Stal, T.~Stefaniak and G.~Weiglein,
  JHEP {\bf 1411}, 039 (2014)
  \doi{10.1007/JHEP11(2014)039}
  [\arxiv{1403.1582} [hep-ph]];
  K.~Cheung, J.~S.~Lee and P.~Y.~Tseng,
  Phys.\ Rev.\ D {\bf 90}, 095009 (2014)
  \doi{10.1103/PhysRevD.90.095009}
  [\arxiv{1407.8236} [hep-ph]];
  J.~B.~Flament,
  \arxiv{1504.07919} [hep-ph];
  B.~Dumont, S.~Fichet and G.~von Gersdorff,
  JHEP {\bf 1307}, 065 (2013)
  \doi{10.1007/JHEP07(2013)065}
  [\arxiv{1304.3369} [hep-ph]];
  J.~Ellis, V.~Sanz and T.~You,
  JHEP {\bf 1407}, 036 (2014)
  \doi{10.1007/JHEP07(2014)036}
  [\arxiv{1404.3667} [hep-ph]];
  S.~Fichet and G.~Moreau,
  Nucl.\ Phys.\ B {\bf 905}, 391 (2016)
  \doi{10.1016/j.nuclphysb.2016.02.019}
  [\arxiv{1509.00472} [hep-ph]];
  G.~Buchalla, O.~Cata, A.~Celis and C.~Krause,
  Eur.\ Phys.\ J.\ C {\bf 76}, no. 5, 233 (2016)
  \doi{10.1140/epjc/s10052-016-4086-9}
  [\arxiv{1511.00988} [hep-ph]];
  S.~Banerjee, S.~Mukhopadhyay and B.~Mukhopadhyaya,
  Phys.\ Rev.\ D {\bf 89}, no. 5, 053010 (2014)
  \doi{10.1103/PhysRevD.89.053010}
  [\arxiv{1308.4860} [hep-ph]];
  L.~Bian, J.~Shu and Y.~Zhang,
  JHEP {\bf 1509}, 206 (2015)
  \doi{10.1007/JHEP09(2015)206}
  [\arxiv{1507.02238} [hep-ph]].

\bibitem{runI_ex}
  G.~Aad {\it et al.} [ATLAS Collaboration],
  Phys.\ Lett.\ B {\bf 726}, 88 (2013)
  \doi{10.1016/j.physletb.2014.05.011};
  [\arxiv{1307.1427} [hep-ex]];
  S.~Chatrchyan {\it et al.} [CMS Collaboration],
  JHEP {\bf 1306}, 081 (2013)
  \doi{10.1007/JHEP06(2013)081}
  [\arxiv{1303.4571} [hep-ex]].
  
\bibitem{laura} 
  L.~Reina, J.~de Blas, M.~Ciuchini, E.~Franco, D.~Ghosh, S.~Mishima, M.~Pierini and L.~Silvestrini,
  PoS EPS {\bf -HEP2015}, 187 (2015)
  \doi{10.22323/1.234.0187}.

\bibitem{runII_eng} 
  J.~Ellis, C.~W.~Murphy, V.~Sanz and T.~You,
  JHEP {\bf 1806}, 146 (2018)
  \doi{10.1007/JHEP06(2018)146}
  [\arxiv{1803.03252} [hep-ph]].

\bibitem{runII_concha} 
  E.~d.~S.~Almeida, A.~Alves, N.~R.~Agostinho, O.~J.~P.~Eboli and M.~C.~Gonzalez-Garcia,
  \arxiv{1812.01009} [hep-ph].

\bibitem{validity} 
  For a comprehensive discussion on the validity of truncated 
  effective Lagrangians see \eg 
  A.~Biek\"otter, J.~Brehmer and T.~Plehn,
  Phys.\ Rev.\ D {\bf 94}, no. 5, 055032 (2016)
  \doi{10.1103/PhysRevD.94.055032}
  [\arxiv{1602.05202} [hep-ph]];
  R.~Contino, A.~Falkowski, F.~Goertz, C.~Grojean and F.~Riva,
  JHEP {\bf 1607}, 144 (2016)
  \doi{10.1007/JHEP07(2016)144}
  [\arxiv{1604.06444} [hep-ph]].

\bibitem{englert}
  C.~Englert, R.~Kogler, H.~Schulz and M.~Spannowsky,
  Eur.\ Phys.\ J.\ C {\bf 76}, no. 7, 393 (2016)
  \doi{10.1140/epjc/s10052-016-4227-1}
  [\arxiv{1511.05170} [hep-ph]].

\bibitem{eft_model}
  See \eg
  J.~Brehmer, A.~Freitas, D.~Lopez-Val and T.~Plehn,
  Phys.\ Rev.\ D {\bf 93}, no. 7, 075014 (2016)
  \doi{10.1103/PhysRevD.93.075014}
  [\arxiv{1510.03443} [hep-ph]];
  A.~Biek\"otter, A.~Knochel, M.~Kr\"amer, D.~Liu and F.~Riva,
  Phys.\ Rev.\ D {\bf 91}, 055029 (2015)
  \doi{10.1103/PhysRevD.91.055029}
  [\arxiv{1406.7320} [hep-ph]];
  S.~Dawson, I.~M.~Lewis and M.~Zeng,
  Phys.\ Rev.\ D {\bf 91}, 074012 (2015)
  \doi{10.1103/PhysRevD.91.074012}
  [\arxiv{1501.04103} [hep-ph]];
  M.~Gorbahn, J.~M.~No and V.~Sanz,
  JHEP {\bf 1510}, 036 (2015)
  \doi{10.1007/JHEP10(2015)036}
  [\arxiv{1502.07352} [hep-ph]];
  A.~Drozd, J.~Ellis, J.~Quevillon and T.~You,
  JHEP {\bf 1506}, 028 (2015)
  \doi{10.1007/JHEP06(2015)028}
  [\arxiv{1504.02409} [hep-ph]];
  A.~Freitas, D.~Lopez-Val and T.~Plehn,
  Phys.\ Rev.\ D {\bf 94}, no. 9, 095007 (2016)
  \doi{10.1103/PhysRevD.94.095007}
  [\arxiv{1607.08251} [hep-ph]].
  
 \bibitem{higgsmultiplets}
  F.~Bonnet, M.~B.~Gavela, T.~Ota and W.~Winter,
  Phys.\ Rev.\ D {\bf 85}, 035016 (2012)
  \doi{10.1103/PhysRevD.85.035016}
  [\arxiv{1105.5140} [hep-ph]];
  F.~Bonnet, T.~Ota, M.~Rauch and W.~Winter,
  Phys.\ Rev.\ D {\bf 86}, 093014 (2012)
  \doi{10.1103/PhysRevD.86.093014}
  [\arxiv{1207.4599} [hep-ph]].

\bibitem{qcd_eft}
  E.~H.~Simmons,
  Phys.\ Lett.\ B {\bf 226}, 132 (1989)
  \doi{10.1016/0370-2693(89)90301-8};
  H.~K.~Dreiner, A.~Duff and D.~Zeppenfeld,
  Phys.\ Lett.\ B {\bf 282}, 441 (1992)
  \doi{10.1016/0370-2693(92)90666-R};
  L.~J.~Dixon and Y.~Shadmi,
  Nucl.\ Phys.\ B {\bf 423}, 3 (1994)
  \doi{10.1016/0550-3213(94)90563-0} 
  [\arxiv{hep-ph/9312363}];
  P.~L.~Cho and E.~H.~Simmons,
  Phys.\ Rev.\ D {\bf 51}, 2360 (1995)
  \doi{10.1103/PhysRevD.51.2360}
  [\arxiv{hep-ph/9408206}];
  S.~Alioli, M.~Farina, D.~Pappadopulo and J.~T.~Ruderman,
  JHEP {\bf 1707}, 097 (2017)
  \doi{10.1007/JHEP07(2017)097}
  [\arxiv{1706.03068} [hep-ph]];
  V.~Hirschi, F.~Maltoni, I.~Tsinikos and E.~Vryonidou,
  JHEP {\bf 1807}, 093 (2018)
  \doi{10.1007/JHEP07(2018)093}
  [\arxiv{1806.04696} [hep-ph]].

\bibitem{krauss}
  F.~Krauss, S.~Kuttimalai and T.~Plehn,
  Phys.\ Rev.\ D {\bf 95}, no. 3, 035024 (2017)
  \doi{10.1103/PhysRevD.95.035024}
  [\arxiv{1611.00767} [hep-ph]].

\bibitem{top_eft}
  C.~Zhang and S.~Willenbrock,
  Phys.\ Rev.\ D {\bf 83}, 034006 (2011)
  \doi{10.1103/PhysRevD.83.034006}
  [\arxiv{1008.3869} [hep-ph]];
  C.~Zhang,
  Chin.\ Phys.\ C {\bf 42}, no. 2, 023104 (2018)
  \doi{10.1088/1674-1137/42/2/023104}
  [\arxiv{1708.05928} [hep-ph]];
  C.~Degrande, F.~Maltoni, K.~Mimasu, E.~Vryonidou and C.~Zhang,
  JHEP {\bf 1810}, 005 (2018)
  \doi{10.1007/JHEP10(2018)005}
  [\arxiv{1804.07773} [hep-ph]];
  M.~Chala, J.~Santiago and M.~Spannowsky,
  \arxiv{1809.09624} [hep-ph].

\bibitem{topfitter}
  A.~Buckley, C.~Englert, J.~Ferrando, D.~J.~Miller, L.~Moore, M.~Russell and C.~D.~White,
  JHEP {\bf 1604}, 015 (2016)
  \doi{10.1007/JHEP04(2016)015}
  [\arxiv{1512.03360} [hep-ph]].

\bibitem{flavor_eft} 
  see \eg
  J.~Aebischer, J.~Kumar, P.~Stangl and D.~M.~Straub,
  \arxiv{1810.07698} [hep-ph].

\bibitem{lep_gauge} 
  E.~Masso and V.~Sanz,
  Phys.\ Rev.\ D {\bf 87}, no. 3, 033001 (2013)
  \doi{10.1103/PhysRevD.87.033001}
  [\arxiv{1211.1320} [hep-ph]];
  A.~Falkowski, M.~Gonzalez-Alonso, A.~Greljo and D.~Marzocca,
  Phys.\ Rev.\ Lett.\  {\bf 116}, no. 1, 011801 (2016)
  \doi{10.1103/PhysRevLett.116.011801}
  [\arxiv{1508.00581} [hep-ph]];
  G.~Brooijmans {\it et al.},
  \arxiv{1405.1617} [hep-ph];
  M.~Trott,
  JHEP {\bf 1502}, 046 (2015)
  \doi{10.1007/JHEP02(2015)046}
  [\arxiv{1409.7605} [hep-ph]];
  A.~Falkowski and F.~Riva,
  JHEP {\bf 1502}, 039 (2015)
  \doi{10.1007/JHEP02(2015)039}
  [\arxiv{1411.0669} [hep-ph]];

\bibitem{lhc_gauge}
  T.~Corbett, O.~J.~P.~Eboli, J.~Gonzalez-Fraile and M.~C.~Gonzalez-Garcia,
  Phys.\ Rev.\ Lett.\  {\bf 111}, 011801 (2013)
  \doi{10.1103/PhysRevLett.111.011801}
  [\arxiv{1304.1151} [hep-ph]];
  J.~Ellis, V.~Sanz and T.~You,
  JHEP {\bf 1503}, 157 (2015)
  \doi{10.1007/JHEP03(2015)157}
  [\arxiv{1410.7703} [hep-ph]];
  A.~Falkowski, M.~Gonzalez-Alonso, A.~Greljo, D.~Marzocca and M.~Son,
  JHEP {\bf 1702}, 115 (2017)
  \doi{10.1007/JHEP02(2017)115}
  [\arxiv{1609.06312} [hep-ph]];
  A.~Azatov, J.~Elias-Miro, Y.~Reyimuaji and E.~Venturini,
  JHEP {\bf 1710}, 027 (2017)
  \doi{10.1007/JHEP10(2017)027}
  [\arxiv{1707.08060} [hep-ph]];
  R.~Franceschini, G.~Panico, A.~Pomarol, F.~Riva and A.~Wulzer,
  JHEP {\bf 1802}, 111 (2018)
  \doi{10.1007/JHEP02(2018)111}
  [\arxiv{1712.01310} [hep-ph]];
  D.~Liu and L.~T.~Wang,
  \arxiv{1804.08688} [hep-ph].

\bibitem{lep_lhc} 
  M.~Farina, G.~Panico, D.~Pappadopulo, J.~T.~Ruderman, R.~Torre and A.~Wulzer,
  Phys.\ Lett.\ B {\bf 772}, 210 (2017)
  \doi{10.1016/j.physletb.2017.06.043}
  [\arxiv{1609.08157} [hep-ph]];
  C.~Grojean, M.~Montull and M.~Riembau,
  \arxiv{1810.05149} [hep-ph].

\bibitem{fermionic}
  Z.~Zhang,
  Phys.\ Rev.\ Lett.\  {\bf 118}, no. 1, 011803 (2017)
  \doi{10.1103/PhysRevLett.118.011803}
  [\arxiv{1610.01618} [hep-ph]];
  T.~Corbett, O.~J.~P.~Eboli and M.~C.~Gonzalez-Garcia,
  Phys.\ Rev.\ D {\bf 96}, no. 3, 035006 (2017)
  \doi{10.1103/PhysRevD.96.035006}
  [\arxiv{1705.09294} [hep-ph]];
  J.~Baglio, S.~Dawson and I.~M.~Lewis,
  Phys.\ Rev.\ D {\bf 96}, no. 7, 073003 (2017)
  \doi{10.1103/PhysRevD.96.073003}
  [\arxiv{1708.03332} [hep-ph]];
  J.~Baglio, S.~Dawson and I.~M.~Lewis,
  \arxiv{1812.00214} [hep-ph];
  A.~Alves, N.~Rosa-Agostinho, O.~J.~P.~Eboli and M.~C.~Gonzalez--Garcia,
  Phys.\ Rev.\ D {\bf 98}, no. 1, 013006 (2018)
  \doi{10.1103/PhysRevD.98.013006}
  [\arxiv{1805.11108} [hep-ph]];
  S.~Dawson and A.~Ismail,
  Phys.\ Rev.\ D {\bf 98}, no. 9, 093003 (2018)
  \doi{10.1103/PhysRevD.98.093003}
  [\arxiv{1808.05948} [hep-ph]].

\bibitem{sfitter_orig}
  R.~Lafaye, T.~Plehn, M.~Rauch, D.~Zerwas and M.~Duhrssen,
  JHEP {\bf 0908}, 009 (2009)
  \doi{10.1088/1126-6708/2009/08/009}
  [\arxiv{0904.3866} [hep-ph]].

\bibitem{sfitter_delta}
  M.~Klute, R.~Lafaye, T.~Plehn, M.~Rauch and D.~Zerwas,
  Phys.\ Rev.\ Lett.\  {\bf 109}, 101801 (2012)
  \doi{10.1103/PhysRevLett.109.101801}
  [\arxiv{1205.2699} [hep-ph]];
  T.~Plehn and M.~Rauch,
  EPL {\bf 100}, no. 1, 11002 (2012)
  \doi{10.1209/0295-5075/100/11002}
  [\arxiv{1207.6108} [hep-ph]].

\bibitem{cp}
  J.~Brehmer, F.~Kling, T.~Plehn and T.~M.~P.~Tait,
  Phys.\ Rev.\ D {\bf 97}, no. 9, 095017 (2018)
  \doi{10.1103/PhysRevD.97.095017}
  [\arxiv{1712.02350} [hep-ph]];
  and not quite convincingly pretended in
  F.~U.~Bernlochner, C.~Englert, C.~Hays, K.~Lohwasser, H.~Mildner, A.~Pilkington, D.~D.~Price and M.~Spannowsky,
  \arxiv{1808.06577} [hep-ph].

\bibitem{hhh}
  For a meaningful measurement at future colliders see \eg
  S.~Di Vita, C.~Grojean, G.~Panico, M.~Riembau and T.~Vantalon,
  JHEP {\bf 1709}, 069 (2017)
  \doi{10.1007/JHEP09(2017)069}
  [\arxiv{1704.01953} [hep-ph]];
  D.~Goncalves, T.~Han, F.~Kling, T.~Plehn and M.~Takeuchi,
  Phys.\ Rev.\ D {\bf 97}, no. 11, 113004 (2018)
  \doi{10.1103/PhysRevD.97.113004}
  [\arxiv{1802.04319} [hep-ph]];
  J.~Chang, K.~Cheung, J.~S.~Lee, C.~T.~Lu and J.~Park,
  \arxiv{1804.07130} [hep-ph];
  S.~Homiller and P.~Meade,
  \arxiv{1811.02572} [hep-ph];
  A.~Biek\"otter, D.~Goncalves, T.~Plehn, M.~Takeuchi and D.~Zerwas,
  \arxiv{1811.08401} [hep-ph];
  S.~Borowka, C.~Duhr, F.~Maltoni, D.~Pagani, A.~Shivaji and X.~Zhao,
  \arxiv{1811.12366} [hep-ph].
  
\bibitem{deBlas:2018tjm} 
  J.~de Blas, O.~Eberhardt and C.~Krause,
  JHEP {\bf 1807}, 048 (2018)
  doi:10.1007/JHEP07(2018)048
  [arXiv:1803.00939 [hep-ph]].

\bibitem{eboli_zeppenfeld}
  O.~J.~P.~Eboli and D.~Zeppenfeld,
  Phys.\ Lett.\ B {\bf 495}, 147 (2000)
  \doi{ 10.1016/S0370-2693(00)01213-2}
  [\arxiv{hep-ph/0009158}];
  C.~Bernaciak, T.~Plehn, P.~Schichtel and J.~Tattersall,
  Phys.\ Rev.\ D {\bf 91}, 035024 (2015)
  \doi{10.1103/PhysRevD.91.035024}
  [\arxiv{1411.7699} [hep-ph]]l;
  for a comprehensive comparison of different signatures see
  A.~Biek\"otter, F.~Keilbach, R.~Moutafis, T.~Plehn and J.~Thompson,
  SciPost Phys.\  {\bf 4}, no. 6, 035 (2018)
  \doi{10.21468/SciPostPhys.4.6.035}
  [\arxiv{1712.03973} [hep-ph]].



\bibitem{kaoru_tgv}
  K.~Hagiwara, R.~D.~Peccei, D.~Zeppenfeld and K.~Hikasa,
  Nucl.\ Phys.\ B {\bf 282}, 253 (1987)
  \doi{10.1016/0550-3213(87)90685-7}.

\bibitem{nonlinear1}
  R.~Alonso, M.~B.~Gavela, L.~Merlo, S.~Rigolin and J.~Yepes,
  Phys.\ Lett.\ B {\bf 722}, 330 (2013)
  \doi{10.1016/j.physletb.2013.04.037} 
  [\arxiv{1212.3305} [hep-ph]];
  G.~Buchalla, O.~Cata and C.~Krause,
  Nucl.\ Phys.\ B {\bf 880}, 552 (2014)
  \doi{10.1016/j.nuclphysb.2016.09.010} 
  [\arxiv{1307.5017} [hep-ph]];
  M.~B.~Gavela, J.~Gonzalez-Fraile, M.~C.~Gonzalez-Garcia, L.~Merlo, S.~Rigolin and J.~Yepes,
  JHEP {\bf 1410}, 044 (2014)
  \doi{10.1007/JHEP10(2014)044}
  [\arxiv{1406.6367} [hep-ph]].

\bibitem{nonlinear2}  
  I.~Brivio, J.~Gonzalez-Fraile, M.~C.~Gonzalez-Garcia and L.~Merlo,
  Eur.\ Phys.\ J.\ C {\bf 76}, no. 7, 416 (2016)
  \doi{10.1140/epjc/s10052-016-4211-9}
  [\arxiv{1604.06801} [hep-ph]].
  

\bibitem{Alioli:2017ces} 
  S.~Alioli, V.~Cirigliano, W.~Dekens, J.~de Vries and E.~Mereghetti,
  JHEP {\bf 1705}, 086 (2017)
  doi:10.1007/JHEP05(2017)086
  [arXiv:1703.04751 [hep-ph]].
  
\bibitem{Cirigliano:2009wk} 
  V.~Cirigliano, J.~Jenkins and M.~Gonzalez-Alonso,
  Nucl.\ Phys.\ B {\bf 830}, 95 (2010)
  doi:10.1016/j.nuclphysb.2009.12.020
  [arXiv:0908.1754 [hep-ph]].

\bibitem{Falkowski:2017pss} 
  A.~Falkowski, M.~González-Alonso and K.~Mimouni,
  JHEP {\bf 1708}, 123 (2017)
  doi:10.1007/JHEP08(2017)123
  [arXiv:1706.03783 [hep-ph]].
  

\bibitem{unitarity} 
  T.~Corbett, O.~J.~P.~Eboli and M.~C.~Gonzalez-Garcia,
  Phys.\ Rev.\ D {\bf 96}, no. 3, 035006 (2017)
  \doi{10.1103/PhysRevD.96.035006}
  [\arxiv{1705.09294} [hep-ph]].

\bibitem{ewwg} 
  S.~Schael {\it et al.} [ALEPH and DELPHI and L3 and OPAL and SLD Collaborations and LEP Electroweak Working Group and SLD Electroweak Group and SLD Heavy Flavour Group],
  Phys.\ Rept.\  {\bf 427}, 257 (2006)
  \doi{10.1016/j.physrep.2005.12.006}
  [hep-ex/0509008].
  
\bibitem{pdg}
  M.~Tanabashi {\it et al.} [Particle Data Group],
  Phys.\ Rev.\ D {\bf 98}, no. 3, 030001 (2018)
  \doi{10.1103/PhysRevD.98.030001}
  
\bibitem{deBlas:2017wmn} 
  J.~de Blas, M.~Ciuchini, E.~Franco, S.~Mishima, M.~Pierini, L.~Reina and L.~Silvestrini,
  PoS EPS {\bf -HEP2017}, 467 (2017)
  \doi{10.22323/1.314.0467}
  [\arxiv{1710.05402} [hep-ph]].
  
\bibitem{eft_higgs_top}
  F.~Maltoni, E.~Vryonidou and C.~Zhang,
  JHEP {\bf 1610}, 123 (2016)
  \doi{10.1007/JHEP10(2016)123}
  [\arxiv{1607.05330} [hep-ph]];
  E.~Vryonidou and C.~Zhang,
  JHEP {\bf 1808}, 036 (2018)
  \doi{10.1007/JHEP08(2018)036}
  [\arxiv{1804.09766} [hep-ph]].

\bibitem{Sirunyan:2018ygk} 
  A.~M.~Sirunyan {\it et al.} [CMS Collaboration],
  JHEP {\bf 1806}, 101 (2018)
  \doi{10.1007/JHEP06(2018)101}
  [\arxiv{1803.06986} [hep-ex]].

\bibitem{madgraph}
  J.~Alwall {\it et al.},
  JHEP {\bf 1407}, 079 (2014)
  \doi{10.1007/JHEP07(2014)079}
  [\arxiv{1405.0301} [hep-ph]].

\bibitem{pythia}
  T.~Sj\"ostrand {\it et al.},
  Comput.\ Phys.\ Commun.\  {\bf 191}, 159 (2015)
  \doi{10.1016/j.cpc.2015.01.024}
  [\arxiv{1410.3012} [hep-ph]].

\bibitem{delphes} 
  J.~de Favereau {\it et al.} [DELPHES 3 Collaboration],
  JHEP {\bf 1402}, 057 (2014)
  \doi{10.1007/JHEP02(2014)057}
  [\arxiv{1307.6346} [hep-ex]].

\bibitem{ATLAS13WW} 
  M.~Aaboud {\it et al.} [ATLAS Collaboration],
  ATLAS-CONF-2018-004.

\bibitem{ATLAS13tthLep} 
  M.~Aaboud {\it et al.} [ATLAS Collaboration],
  Phys.\ Rev.\ D {\bf 97}, no. 7, 072003 (2018)
  \doi{10.1103/PhysRevD.97.072003}
  [\arxiv{1712.08891} [hep-ex]].

\bibitem{CMS13WW} 
  A.~M.~Sirunyan {\it et al.} [CMS Collaboration],
  [\arxiv{1806.05246} [hep-ex]].
  
\bibitem{CMS13tthLep1} 
  A.~M.~Sirunyan {\it et al.} [CMS Collaboration],
  CMS-PAS-HIG-17-004.

\bibitem{CMS13tthLep2} 
  A.~M.~Sirunyan {\it et al.} [CMS Collaboration],
  \arxiv{1803.05485} [hep-ex].

\bibitem{ATLAS13ZZ} 
  M.~Aaboud {\it et al.} [ATLAS Collaboration],
  JHEP {\bf 1803}, 095 (2018)
  \doi{10.1007/JHEP03(2018)095}
  \arxiv{1712.02304} [hep-ex].
  
\bibitem{CMS13ZZ} 
  A.~M.~Sirunyan {\it et al.} [CMS Collaboration],
  JHEP {\bf 1711}, 047 (2017)
  \doi{10.1007/JHEP11(2017)047}
  \arxiv{1706.09936} [hep-ex].

\bibitem{CMS13ZZv2} 
  A.~M.~Sirunyan {\it et al.} [CMS Collaboration],
  CMS-PAS-HIG-18-001.
  
\bibitem{ATLAS13aa} 
  M.~Aaboud {\it et al.} [ATLAS Collaboration],
  \arxiv{1802.04146} [hep-ex].
  
 \bibitem{CMS13aa} 
  A.~M.~Sirunyan {\it et al.} [CMS Collaboration],
  \arxiv{1804.02716} [hep-ex].
  
\bibitem{CMS13tautau} 
  A.~M.~Sirunyan {\it et al.} [CMS Collaboration],
  Phys.\ Lett.\ B {\bf 779}, 283 (2018)
  \doi{10.1016/j.physletb.2018.02.004}
  \arxiv{1708.00373} [hep-ex].

\bibitem{ATLAS13Za} 
  M.~Aaboud {\it et al.} [ATLAS Collaboration],
  JHEP {\bf 1710}, 112 (2017)
  \doi{10.1007/JHEP10(2017)112}
  \arxiv{1708.00212} [hep-ex].
  
\bibitem{CMS13Za} 
  A.~M.~Sirunyan {\it et al.} [CMS Collaboration],
  \arxiv{1806.05996} [hep-ex].

\bibitem{CMS13WBFInv} 
  A.~M.~Sirunyan [CMS Collaboration],
  CMS-PAS-HIG-17-023.

\bibitem{ATLAS13Vhbb} 
  M.~Aaboud {\it et al.} [ATLAS Collaboration],
  JHEP {\bf 1712}, 024 (2017)
  \doi{10.1007/JHEP12(2017)024}
  \arxiv{1708.03299} [hep-ex].

\bibitem{CMS13Vhbb} 
  A.~M.~Sirunyan {\it et al.} [CMS Collaboration],
  Phys.\ Lett.\ B {\bf 780}, 501 (2018)
  \doi{10.1016/j.physletb.2018.02.050}
 [\arxiv{1709.07497} [hep-ex].

\bibitem{CMS13Vhtautau} 
  A.~M.~Sirunyan [CMS Collaboration],
  CMS-PAS-HIG-18-007.

\bibitem{ATLAS13ZhInv} 
  M.~Aaboud {\it et al.} [ATLAS Collaboration],
  Phys.\ Lett.\ B {\bf 776}, 318 (2018)
  \doi{10.1016/j.physletb.2017.11.049}
  \arxiv{1708.09624} [hep-ex].

\bibitem{CMS13ZhInv} 
  A.~M.~Sirunyan {\it et al.} [CMS Collaboration],
  Eur.\ Phys.\ J.\ C {\bf 78}, no. 4, 291 (2018)
  \doi{10.1140/epjc/s10052-018-5740-1}
  [\arxiv{1711.00431} [hep-ex]].

\bibitem{ATLAS13VhEXO} 
  M.~Aaboud {\it et al.} [ATLAS Collaboration],
  JHEP {\bf 1803}, 174 (2018)
  \doi{10.1007/JHEP03(2018)174}
  \arxiv{1712.06518} [hep-ex].

\bibitem{ATLAS13tthObs} 
  M.~Aaboud {\it et al.} [ATLAS Collaboration],
  Phys.\ Lett.\ B {\bf 784}, 173 (2018)
  \doi{10.1016/j.physletb.2018.07.035}
  \arxiv{1806.00425} [hep-ex].
  
\bibitem{ATLAS13tthbb} 
  M.~Aaboud {\it et al.} [ATLAS Collaboration],
  Phys.\ Rev.\ D {\bf 97}, no. 7, 072016 (2018)
  \doi{10.1103/PhysRevD.97.072016}
  \arxiv{1712.08895} [hep-ex].
  
\bibitem{CMS13tthbb} 
  A.~M.~Sirunyan {\it et al.} [CMS Collaboration],
  \arxiv{1804.03682} [hep-ex].

\bibitem{hdecay}
  R.~Contino, M.~Ghezzi, C.~Grojean, M.~M\"uhlleitner and M.~Spira,
  Comput.\ Phys.\ Commun.\  {\bf 185}, 3412 (2014)
  \doi{10.1016/j.cpc.2014.06.028}
  [\arxiv{1403.3381} [hep-ph]].

\bibitem{feynrules}
  N.~D.~Christensen and C.~Duhr,
  Comput.\ Phys.\ Commun.\  {\bf 180}, 1614 (2009)
  \doi{10.1016/j.cpc.2009.02.018}
  [\arxiv{0806.4194} [hep-ph]].

\bibitem{sally}
  S.~Dawson, private communication.

\bibitem{dennerNLO} 
  M.~Chiesa, A.~Denner and J.~N.~Lang,
  Eur.\ Phys.\ J.\ C {\bf 78}, no. 6, 467 (2018)
  doi:10.1140/epjc/s10052-018-5949-z
  [arXiv:1804.01477 [hep-ph]].

\bibitem{info_geo}
  J.~Brehmer, K.~Cranmer, F.~Kling and T.~Plehn,
  Phys.\ Rev.\ D {\bf 95}, no. 7, 073002 (2017)
  \doi{10.1103/PhysRevD.95.073002}
  [\arxiv{1612.05261} [hep-ph]].

\bibitem{atlas8ww}
  G.~Aad {\it et al.} [ATLAS Collaboration],
  JHEP {\bf 1609}, 029 (2016)
  \doi{10.1007/JHEP09(2016)029}
  [\arxiv{1603.01702} [hep-ex]].

\bibitem{cms8ww}
  V.~Khachatryan {\it et al.} [CMS Collaboration],
  Eur.\ Phys.\ J.\ C {\bf 76}, no. 7, 401 (2016)
  \doi{10.1140/epjc/s10052-016-4219-1}
  [\arxiv{1507.03268} [hep-ex]].

\bibitem{atlas8wz}
  G.~Aad {\it et al.} [ATLAS Collaboration],
  Phys.\ Rev.\ D {\bf 93}, no. 9, 092004 (2016)
  \doi{10.1103/PhysRevD.93.092004}
  [\arxiv{1603.02151} [hep-ex]].

\bibitem{cms78wz} 
  CMS Collaboration [CMS Collaboration],
  CMS-PAS-SMP-12-006.

\bibitem{ATLAS13WZ} 
  M.~Aaboud {\it et al.} [ATLAS Collaboration],
  ATLAS-CONF-2018-034.


\bibitem{interference} 
  A.~Helset and M.~Trott,
  JHEP {\bf 1804}, 038 (2018)
  \doi{10.1007/JHEP04(2018)038}
  [\arxiv{1711.07954} [hep-ph]].

\bibitem{dim8} 
  C.~Hays, A.~Martin, V.~Sanz and J.~Setford,
  \arxiv{1808.00442} [hep-ph].
  
\bibitem{ATLASttXS} 
  The ATLAS collaboration,
  ATLAS-CONF-2015-049.
    
\bibitem{ttheventuallybeatstt} 
  F.~Maltoni, E.~Vryonidou and C.~Zhang,
  JHEP {\bf 1610}, 123 (2016)
  doi:10.1007/JHEP10(2016)123
  [arXiv:1607.05330 [hep-ph]].

\end{thebibliography}
\end{document}